\documentclass[submitting]{cpc}

\usepackage{subfigure,dcolumn}
\usepackage{epstopdf}
\usepackage{mhchem}
\usepackage{upgreek}
\usepackage{float}
\usepackage{graphicx}

\begin{document}

\title{Projected Sensitivity to Atmospheric Neutrino Oscillations using a 1725 m$^{3}$ Liquid-Nitrogen Detector at CJPL}

\thanks{ This work was supported by the National Key Research and Development Program of China (Grant No. 2023YFA1607103) and the National Natural Science Foundation of China (Grants No. 12441512, 11975159, and 11975162).}

\author{Xiao-Yu Peng}
\affiliation{College of Physics, Sichuan University, Chengdu 610065, China}
\author{Shin-Ted Lin}
\email[E-mail: ]{stlin@scu.edu.cn}
\affiliation{College of Physics, Sichuan University, Chengdu 610065, China}
\author{Shu-Kui Liu}
\email[E-mail: ]{liusk@scu.edu.cn}
\affiliation{College of Physics, Sichuan University, Chengdu 610065, China}
\author{Hao-Yang Xing}
\affiliation{College of Physics, Sichuan University, Chengdu 610065, China}
\author{Jing-Jun Zhu}
\affiliation{College of Physics, Sichuan University, Chengdu 610065, China}
\author{Han-Yu Li}
\affiliation{College of Physics, Sichuan University, Chengdu 610065, China}
\author{Bi-Jun Xiao}
\affiliation{College of Physics, Sichuan University, Chengdu 610065, China}
\author{Li-Tao Yang}
\affiliation{ Key Laboratory of Particle and Radiation Imaging (Ministry of Education) and Department of Engineering Physics, Tsinghua University, Beijing 100084, China}
\author{Qian Yue}
\affiliation{ Key Laboratory of Particle and Radiation Imaging (Ministry of Education) and Department of Engineering Physics, Tsinghua University, Beijing 100084, China}
\author{Muhammed Deniz}
\affiliation{ Department of Physics, Dokuz Eylül University, Buca 35160, İzmir, Turkey}
\begin{abstract}

In atmospheric neutrino analyses, the usable zenith-angle range is generally restricted to upward-going events to suppress the cosmic-ray muon background. Motivated by the exceptionally low underground muon background at the China Jinping Underground Laboratory (CJPL), we investigate whether extending the zenith-angle acceptance can improve the sensitivity of atmospheric neutrino oscillation measurements. An existing 1725 m$^{3}$ liquid-nitrogen volume is adopted as the basis of a simplified detector model, in which the accepted zenith-angle range is extended from $\cos\theta_{\mu}\in[-1,0]$ to $\cos\theta_{\mu}\in[-1,0.3]$, and the projected oscillation sensitivity is evaluated for a 10-year exposure. The atmospheric neutrino flux is estimated by interpolating standard flux predictions as a function of geomagnetic latitude. Neutrino interactions in the surrounding rock and the liquid-nitrogen volume are simulated for neutrino energies of $E\ge0.1$ GeV, followed by secondary-particle transport and geometry-based event selection. For the benchmark oscillation parameters $\Delta$m$^{2}_{32}$ = $2.4\times10^{-3}\ \mathrm{eV}^{2}$ and $\sin^{2}\theta_{23}=0.5$, the neutrino-induced muon flux from the surrounding rock is estimated to be $(3.65 \pm 1.00)\times10^{-13}$ cm$^{-2}$ s$^{-1}$ sr$^{-1}$, corresponding to a muon yield of $(0.13 \pm 0.034)$ day$^{-1}$ in the 1725 m$^{3}$ liquid-nitrogen volume. Three-flavor neutrino oscillations are incorporated into a Poisson-likelihood $\chi^{2}$ analysis to evaluate the sensitivity to the oscillation parameters. Extending the usable zenith-angle range leads to a clear improvement in the projected oscillation sensitivity at the 90\% confidence level over a 10-year exposure. The improvement is most pronounced for $\Delta$m$^{2}_{32}$, while the improvement in the constraint on $\sin^{2}\theta_{23}$ near the best-fit point remains relatively modest. These results demonstrate that the extended zenith-angle coverage enabled by the ultra-low cosmic-ray muon background at CJPL can strengthen the projected constraints on atmospheric neutrino oscillation parameters in a large liquid-nitrogen detector.

\end{abstract}

\keywords{Neutrino detection, atmospheric neutrino oscillation parameters, underground laboratory, muon flux}

\maketitle

\section{Introduction}\label{sec:1}
Neutrinos are electrically neutral leptons in the Standard Model that interact through the weak interaction and appear in three flavor states. A broad range of observations with solar \cite{bib:1,bib:2,bib:3}, atmospheric \cite{bib:4,bib:5,bib:6}, reactor \cite{bib:7,bib:8}, and accelerator neutrinos \cite{bib:8,bib:9} has established that neutrinos are massive and undergo flavor oscillations, providing clear evidence for physics beyond the Standard Model. In the standard three-flavor framework, lepton mixing is described by the Pontecorvo-Maki-Nakagawa-Sakata (PMNS) matrix \cite{bib:10,bib:11}, which is commonly parametrized by three mixing angles ($\theta_{12}$, $\theta_{23}$, $\theta_{13}$), two independent mass-squared splittings ($\Delta$m$^{2}_{21}$, $\Delta$m$^{2}_{32}$), and one CP-violating phase $\delta_{CP}$. Solar and reactor measurements have already placed stringent constraints on $\theta_{12}$, $\theta_{13}$, and $\Delta$m$^{2}_{21}$. By contrast, several issues in the atmospheric sector remain open, including the value of $\theta_{23}$, its octant, the CP-violating phase $\delta_{CP}$, and the sign of $\Delta$m$^{2}_{32}$, which determines the neutrino mass ordering. Long-baseline accelerator experiments such as T2K \cite{bib:12} and NOvA \cite{bib:13} have provided precise constraints on $\theta_{23}$ and \textbar$\Delta$m$^{2}_{32}$\textbar and have substantially improved the sensitivity to $\delta_{CP}$. However, a definitive determination of the $\theta_{23}$ octant, $\delta_{CP}$, and the mass ordering is still lacking.

Atmospheric neutrinos provide a natural and independent probe that complements accelerator-based measurements. They are produced mainly through the decays of pions and kaons generated when high-energy cosmic rays interact with nuclei in the upper atmosphere \cite{bib:14}. Their spectrum covers a broad energy range, with most of the flux distributed from tens of MeV to tens of GeV \cite{bib:15,bib:16,bib:17}. Because their production and propagation are affected by the geomagnetic field, the atmospheric neutrino flux exhibits a pronounced latitude dependence. In addition, the corresponding baselines extend from tens of kilometers to approximately the Earth’s diameter. This wide coverage in both energy and path length makes atmospheric neutrinos particularly sensitive to $\sin^{2}\theta_{23}$ and $\Delta$m$^{2}_{32}$. Neutrinos that traverse the Earth also experience matter effects that depend on the propagation distance and the Earth density profile, and therefore on the zenith-angle. Measurements of event rates as functions of energy and zenith-angle thus provide a direct handle on oscillation physics and offer an important way to constrain the parameters of the three-flavor framework.

Atmospheric neutrino detectors can be broadly classified into two categories: Cherenkov detectors and calorimetric detectors. Cherenkov detectors are commonly realized either as imaging Cherenkov detectors or as Cherenkov telescope arrays, while calorimetric detectors are typically implemented as magnetized solid calorimeters or scintillator-based calorimeters. In all of these approaches, neutrino interactions are inferred from secondary charged leptons, primarily $\mu^{\pm}$ and $e^{\pm}$, produced when neutrinos interact in the detector medium or in the surrounding material. A central experimental difficulty is the cosmic-ray muon background (CMBg), since down-going cosmic-ray muons can produce signals similar to those of neutrino-induced muons. Atmospheric neutrino experiments are therefore usually located deep underground or underwater in order to suppress the cosmic-ray muon flux. The remaining background is further reduced by directional selections, such as requiring upward-going events, and, where applicable, by dedicated muon veto systems.

Among current atmospheric neutrino experiments, Super-Kamiokande operates underground with an overburden of about 2700 m water equivalent (m.w.e.) \cite{bib:18}. Its 22.5 kton fiducial-volume ultra-pure water imaging Cherenkov detector analyzes internal events over nearly the full solid angle and selects external samples mainly in the up-going hemisphere, typically with $\cos\theta_{\mu}\in[-1,0]$ \cite{bib:19}. The IceCube experiment, through its DeepCore sub-array deployed roughly 1200 m below the Antarctic ice surface, uses a dense Cherenkov array with an effective instrumented mass of nearly 10,000 tons and analyzes external-event samples extending to $\cos\theta_{\mu}\in[-1,0.1]$ \cite{bib:20}. Located at a depth of about 2070 m.w.e. \cite{bib:18}, the MINOS+ experiment employs a 5.6 kton magnetized iron tracking calorimeter and selects through-going external events in the range $\cos\theta_{\mu}\in[-1,0.05]$ \cite{bib:21}. KM3NeT-ORCA is a deep-sea Cherenkov telescope array instrumenting roughly 7000 kton of seawater and primarily analyzes external events in the up-going hemisphere, $\cos\theta_{\mu}\in[-1,0]$ \cite{bib:23}. Complementary information comes from long-baseline accelerator experiments such as T2K and NOvA, which probe oscillations along fixed beam directions using a water Cherenkov detector and a liquid-scintillator tracking calorimeter, respectively. These experiments currently provide strong constraints on $\sin^{2}\theta_{23}$ and $\Delta$m$^{2}_{32}$ \cite{bib:12,bib:13}. Taken together, these results show that control of the CMBg and access to a wide zenith-angle range are both crucial for improving atmospheric neutrino sensitivity.

China Jinping Underground Laboratory (CJPL), located in Sichuan Province on the southeastern margin of the Qinghai–Tibet Plateau, is one of the deepest underground research facilities in the world. It is covered by a granite overburden with an average thickness of about 2400 m, corresponding to roughly 6700 m.w.e. \cite{bib:24,bib:25}. This depth leads to an exceptionally low CMBg, with a measured muon flux of 3.53$\times$10$^{-10}$ cm$^{-2}$ s$^{-1}$ \cite{bib:26,bib:27}. Such a low muon background makes CJPL a particularly interesting site for atmospheric neutrino studies. Simulations by Guo et al. \cite{bib:26} and Zhang et al. \cite{bib:27} indicate that the CMBg can be strongly suppressed over the zenith-angle range $\cos\theta_{\mu}\in[-1,0.3]$. In this work, this interval is treated as the usable analysis window for atmospheric neutrino event selection. This makes it possible to include near-horizontal events and may improve the sensitivity to the atmospheric oscillation parameters $\sin^{2}\theta_{23}$ and $\Delta$m$^{2}_{32}$.

A large cylindrical cryostat is available in CJPL Phase II as part of the CDEX experimental infrastructure, with a diameter of 13 m and a height of 13 m \cite{bib:28}. The vessel contains 1725 m$^{3}$ of high-purity liquid nitrogen and is surrounded by rock \cite{bib:29}. High-purity liquid nitrogen has optical properties that make it of interest as a possible Cherenkov medium. Its refractive index is about 1.2, the absorption length for Cherenkov photons is approximately 50 m, and the attenuation length is about 30 m. In addition, recent developments in photomultiplier-based light readout suggest that photon detection in liquid nitrogen at 77 K is technically feasible \cite{bib:30}. Motivated by these considerations, we take this existing liquid-nitrogen volume as the basis of a simplified detection scenario. In the present study, we focus on muon signals, including muons produced by neutrino interactions in the surrounding rock and in the liquid-nitrogen volume.

Motivated by the ultra-low CMBg under CJPL conditions, we carry out a site-specific sensitivity study under a simplified liquid-nitrogen-based detection scenario. The projected sensitivity to the oscillation parameters $\sin^{2}\theta_{23}$ and $\Delta$m$^{2}_{32}$ is evaluated for a 10-year exposure using events in the zenith-angle range $\cos\theta_{\mu}\in[-1,0.3]$. In Sec. \ref{sec:2}, the atmospheric neutrino flux at CJPL is estimated from the geomagnetic-latitude dependence of standard flux predictions.  In Secs.~\ref{sec:3} and~\ref{sec:4}, the NuWro event generator \cite{bib:31} and the Geant4 simulation toolkit \cite{bib:32} are used to obtain the probability migration matrices from the true neutrino energy and angle to the visible energy and angle of the observable muon, the expected rate of neutrino-induced muon events, the event selection strategy, and the detector acceptance. In Sec. \ref{sec:5}, three-flavor oscillations are implemented with Prob3++ \cite{bib:33}, and the oscillation analysis is described, including the simulation workflow, the oscillation framework, and the $\chi^{2}$ method used to derive the projected sensitivities. The analysis is performed in the observable space defined by the visible energy and the muon zenith-angle cosine at the entrance to the liquid-nitrogen volume, rather than by event-by-event reconstruction of the true neutrino energy. Finally, Sec. \ref{sec:6} presents the discussion and conclusions. This study provides a benchmark estimate of site-specific atmospheric neutrino oscillation sensitivity under CJPL conditions and offers a quantitative reference for liquid-nitrogen-based detector scenarios under deep-underground conditions.

\section{Atmospheric neutrino flux under CJPL conditions}\label{sec:2}
High-energy cosmic-ray protons and nuclei interact with nitrogen and oxygen in the atmosphere, producing secondary mesons, mainly pions and kaons. The decays of these mesons generate muons and neutrinos, together with their antiparticles. The muons subsequently decay on microsecond timescales and produce additional neutrinos and charged leptons. At low energies ($\lesssim$ 1 GeV), atmospheric neutrino production is mainly governed by the decay chain shown in Eq.~\eqref{eq1}, whereas at high energies ($\gtrsim$ 10 GeV) the contribution described by Eq.~\eqref{eq2} becomes dominant.

\begin{equation}\label{eq1}
\begin{aligned}
\pi^{\pm} &\longrightarrow \mu^{\pm} + \nu_{\mu}^{(\mp)}, \\
\mu^{\pm} &\longrightarrow e^{\pm} + \nu_{e}^{(\mp)} + \bar{\nu}_{\mu}^{(\pm)} .
\end{aligned}
\end{equation}
\begin{equation}\label{eq2}
\begin{aligned}
K^{\pm} &\longrightarrow \mu^{\pm} + \nu_{\mu}^{(\mp)}, \\
K^{\pm} &\xrightarrow{\mathrm{Ke3}} + \pi^{0} + e^{\pm} + \nu_{e}^{(\mp)}, \\
K^{\pm} &\xrightarrow{\mathrm{K}\mu3} + \pi^{0} + \mu^{\pm} + \nu_{\mu}^{(\mp)}, \\
K_{L}^{0} &\xrightarrow{\mathrm{Ke3}} + \pi^{\pm} + e^{\mp} + \nu_{e}^{(\mp)}, \\
K_{L}^{0} &\xrightarrow{\mathrm{K}\mu3} + \pi^{\pm} + \mu^{\mp} + \nu_{\mu}^{(\mp)} .
\end{aligned}
\end{equation}

\begin{figure*}[!htb]
\includegraphics
  [width=0.8\hsize]
  {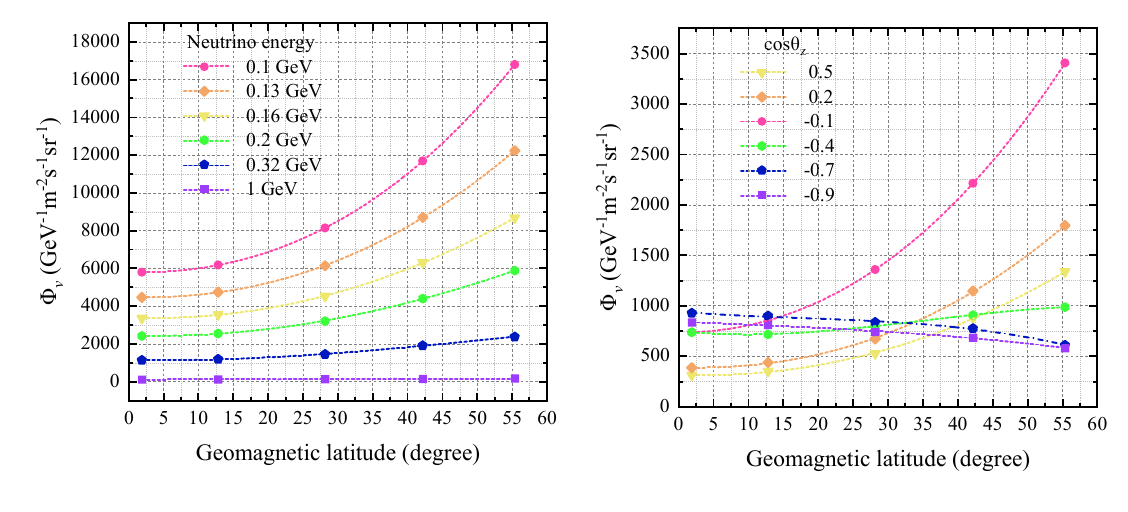}
\caption{Dependence of the unoscillated atmospheric neutrino flux on geomagnetic latitude based on the Honda flux calculations. Left: $\Phi_{\nu}$ as a function of geomagnetic latitude for several selected neutrino energies. Right: $\Phi_{\nu}$ as a function of geomagnetic latitude for several selected zenith-angle intervals. The scatter points correspond to the reference underground laboratories used in the interpolation, and the dashed curves show the fitted latitude dependence.}
\label{fig1}
\end{figure*}

In an atmospheric cascade initiated by primary cosmic rays, the neutrino energy is correlated with the energy of the incident particle. Since the primary cosmic-ray spectrum extends over many orders of magnitude, atmospheric neutrinos also span a broad energy range. The geomagnetic field further modulates the flux by bending the trajectories of charged primaries, with the effect being strongest at low rigidities. Protons with energies below a few GeV are strongly deflected, which suppresses the primary flux at low geomagnetic latitudes and enhances the MeV--GeV atmospheric neutrino flux at higher geomagnetic latitudes. At higher energies ($\ge$ 10 GeV), geomagnetic deflection becomes small, and the atmospheric neutrino flux becomes much less site dependent. As a result, the atmospheric neutrino flux at a given underground laboratory depends strongly on the local geomagnetic conditions. In the present work, we characterize this site dependence by the geomagnetic latitude and use it to estimate the atmospheric neutrino flux at CJPL from established flux calculations for other underground laboratories.

Atmospheric neutrino fluxes in the absence of oscillations have been calculated by Honda et al. \cite{bib:34,bib:35} for a number of underground laboratories worldwide. Table~\ref{tab1} summarizes the geographic latitudes, geomagnetic latitudes, and the corresponding tabulated Honda flux values under solar-minimum conditions for the reference underground laboratories adopted in the present interpolation. Since a dedicated Honda-flux calculation is not currently available for CJPL, we estimate the atmospheric neutrino flux and zenith-angle distribution at CJPL by interpolating the Honda results across laboratories at different geomagnetic latitudes. In this work, the geomagnetic latitude is evaluated in the centered-dipole approximation using the WMM2025 geomagnetic pole. For CJPL, the geographic latitude adopted in the present analysis is 28.15$^{\circ}$, and the corresponding geomagnetic latitude is 18.82$^{\circ}$. These CJPL coordinates are then used to evaluate the fitted latitude dependence and obtain a site-specific estimate of the atmospheric neutrino flux and its zenith-angle distribution at CJPL. The zenith-angle is defined with respect to the local vertical and is expressed as $\cos\theta_{z}$. A neutrino arriving from directly overhead has $\theta_{z}$ = 0$^{\circ}$ ($\cos\theta_{z}$ = 1), a horizontal neutrino has $\theta_{z}$ = 90$^{\circ}$ ($\cos\theta_{z}$ = 0), and an upward-going neutrino arriving from below has $\theta_{z}$ = 180$^{\circ}$ ($\cos\theta_{z}$ = -1). Fig.~\ref{fig1} shows the energy and zenith-angle dependence of the atmospheric neutrino flux at several representative sites. For each $\cos\theta_{z}$ bin and neutrino energy bin, the CJPL flux is obtained by fitting the site dependence of the Honda predictions as a function of geomagnetic latitude and evaluating the fit at the CJPL value. Using this procedure, we obtain the atmospheric neutrino flux at CJPL for $E \ge$ 0.1 GeV, together with its dependence on $\cos\theta_{z}$.

\begin{widetext}
\begin{table*}[!htb]
\centering\caption{Geographic latitudes, geomagnetic latitudes, and the corresponding unoscillated atmospheric neutrino flux values extracted from the Honda solar-minimum tables for the reference underground laboratories adopted in the site-specific interpolation.}
\label{tab1}
\begin{tabular*}{17.5cm} {@{\extracolsep{\fill} } c c c c c c c c}
\toprule
\multicolumn{7}{r}{Neutrino flux (m$^{-2}$ s$^{-1}$ sr$^{-1}$)} \\
\cmidrule(r){4-8}
Laboratory & Geographic latitude ($^{\circ}$) & Geomagnetic latitude ($^{\circ}$) & $\nu_{\mu}$ & $\bar{\nu}_{\mu}$ & $\nu_{e}$ & $\bar{\nu}_{e}$ & Total\\
\midrule
INO      & 9.95  & 1.88  & 912  & 911  & 454  & 409  & 2686 \\
JUNO     & 22.13 & 12.82 & 958  & 953  & 474  & 431  & 2816 \\
Super-K  & 36.43 & 28.16 & 1169 & 1169 & 584  & 531  & 3459 \\
GRN      & 42.45 & 42.18 & 1561 & 1552 & 783  & 705  & 4601 \\
SNO      & 46.47 & 55.37 & 2015 & 2006 & 1043 & 899  & 5963 \\

\bottomrule
\end{tabular*}
\end{table*}
\end{widetext}

\begin{figure*}[!htb]
\includegraphics
  [width=0.8\hsize]
  {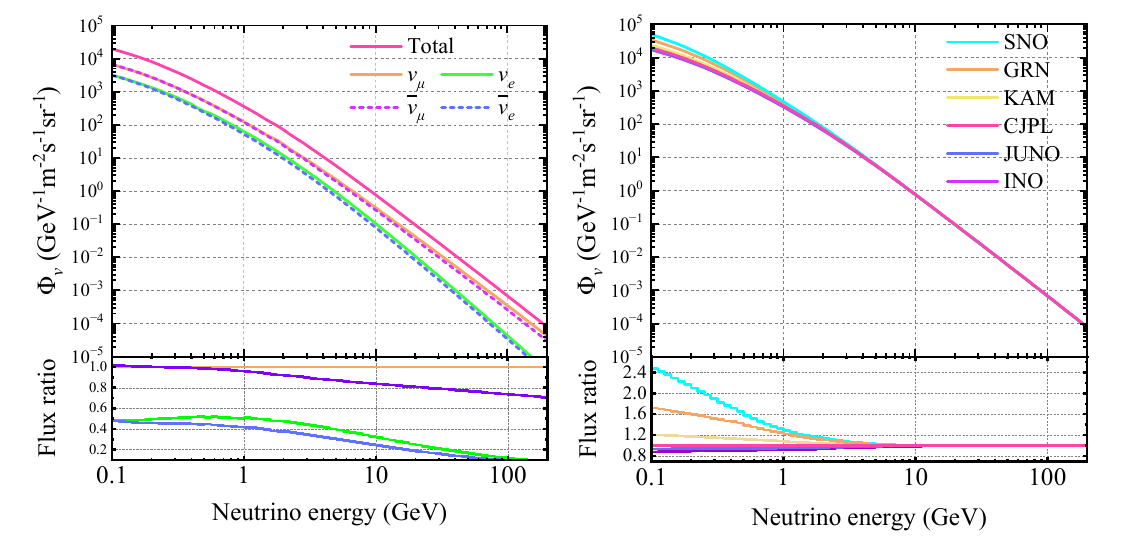}
\caption{Energy dependence of the atmospheric neutrino flux at CJPL obtained from the geomagnetic-latitude interpolation. Left: Energy spectra of muon neutrinos, electron neutrinos, and their antiparticles at CJPL. The flux ratio panel shows each component normalized to the muon neutrino flux. Right: Total flux at CJPL compared with that at other laboratories. The flux ratio panel shows the flux at each laboratory normalized to the total flux at CJPL.}
\label{fig2}
\end{figure*}
\begin{figure*}[!htb]
\includegraphics
  [width=0.8\hsize]
  {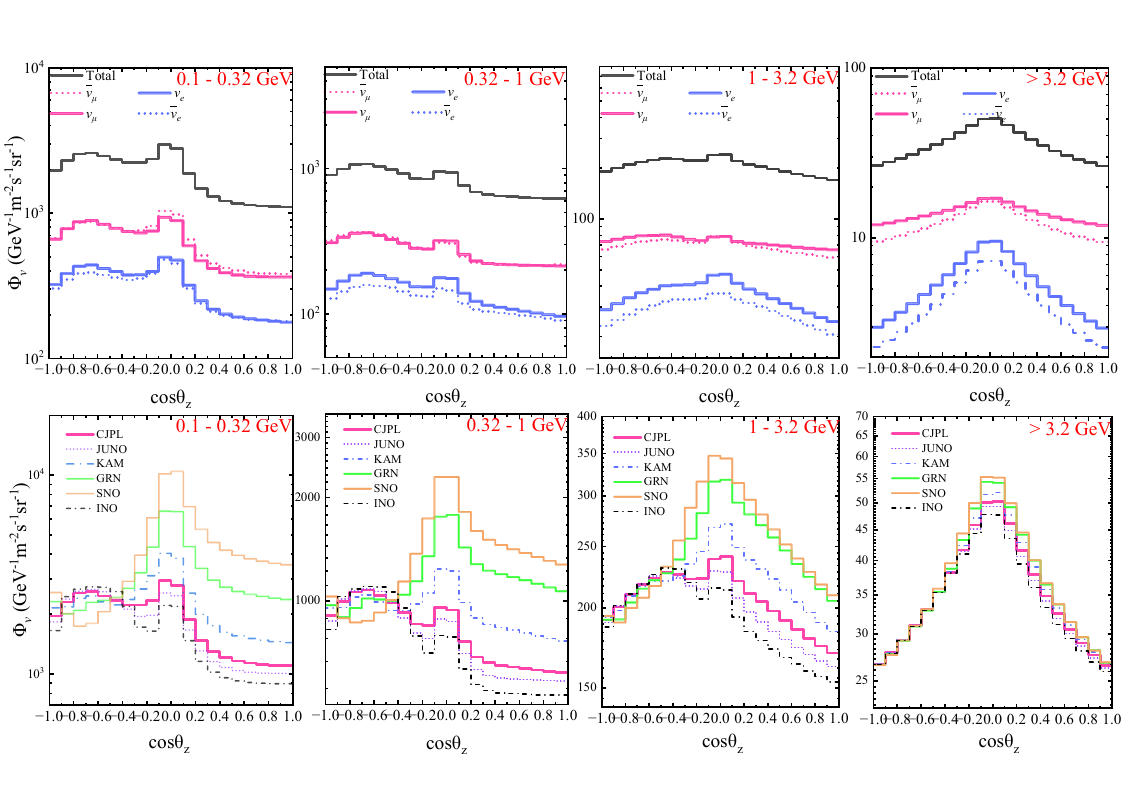}
\caption{Zenith-angle dependence of the CJPL atmospheric neutrino flux inferred from the latitude dependence. From left to right, the panels correspond to 0.1–0.32 GeV, 0.32–1 GeV, 1–3.2 GeV, and $\ge$ 3.2 GeV. The top row shows the CJPL zenith-angle distributions for muon neutrinos, electron neutrinos, and their antiparticles. The bottom row compares the flux as a function of zenith-angle at CJPL with that at other laboratories.}
\label{fig3}
\end{figure*}

Fig.~\ref{fig2} and Fig.~\ref{fig3} show the atmospheric neutrino flux at CJPL obtained from the geomagnetic-latitude interpolation, presented as functions of neutrino energy and zenith-angle, respectively. The left panel of Fig.~\ref{fig2} shows the CJPL flux components for muon neutrinos, antimuon neutrinos, electron neutrinos, and antielectron neutrinos. The right panel of Fig.~\ref{fig2} compares the corresponding energy spectra at different laboratory sites. At sub-GeV to GeV energies, geomagnetic bending of primary cosmic-ray protons leads to pronounced site-to-site variations in the secondary neutrino flux. Above $\sim$10 GeV, geomagnetic effects become weak, and the fluxes at different locations gradually converge.

At low energies, muon neutrinos are produced mainly through pion decay followed by muon decay, whereas electron neutrinos arise predominantly from muon decay. This leads to an approximate 2:1 flavor ratio below the GeV scale. At higher energies, kaon decays become increasingly important, and the relative differences among the four flux components become more pronounced.

Fig.~\ref{fig3} shows the zenith-angle dependence of the atmospheric neutrino flux for several laboratories in different energy intervals. Since the zenith-angle distribution depends on energy, the flux shapes differ among the four ranges 0.1-0.32 GeV, 0.32-1 GeV, 1-3.2 GeV, and $\ge$ 3.2 GeV. Table~\ref{tab2} lists the integrated fluxes of muon neutrinos, electron neutrinos, and their antiparticles over the same energy intervals.

\begin{table}[!htb]
\centering\caption{Differences in neutrino flux within different energy ranges in CJPL.}
\label{tab2}
\begin{tabular*}{8cm} {@{\extracolsep{\fill} } c c c c c  c}
\toprule
\multicolumn{6}{r}{Neutrino flux (m$^{-2}$ s$^{-1}$ sr$^{-1}$)} \\
\cmidrule(r){2-6}
Energy range (GeV)  & $\nu_{\mu}$ & $\bar{\nu}_{\mu}$ & $\nu_{e}$ & $\bar{\nu}_{e}$ & Total\\
\midrule
0.1 - 0.32  & 642  & 656   & 318   & 301   & 1917    \\
0.32 - 1 & 287   & 275   & 142   & 124   & 580    \\
1 - 3.2  &75 & 69   & 35  & 29  & 208    \\
$\ge$3.2  & 14  & 12   & 5  & 4  & 35    \\
$\ge$0.1  &1016  & 1012   & 500  & 458  & 2986   \\

\bottomrule
\end{tabular*}
\end{table}

As shown in Fig.~\ref{fig3}, the zenith-angle distributions of atmospheric neutrinos vary with both site and energy. For neutrino energies below 3.2 GeV, the upward-going flux ($\cos\theta_{z}$ from -1 to 0) exceeds the down-going flux at all sites. At high geomagnetic latitudes, such as GRN and SNO, the distribution is more strongly enhanced toward the horizontal direction. By contrast, at lower-latitude sites, including INO, JUNO, CJPL, and Super-K, the flux tends to peak at intermediate zenith-angles, with prominent contributions around $\cos\theta_{z}\approx -0.7$ and $\cos\theta_{z}\approx 0$. For energies above 3.2 GeV, the latitude dependence becomes much weaker: the fluxes at different locations become increasingly similar, and the upward-going and down-going components approach an approximately symmetric distribution.

\begin{figure}[!htb]
\includegraphics
  [width=0.8\hsize]
  {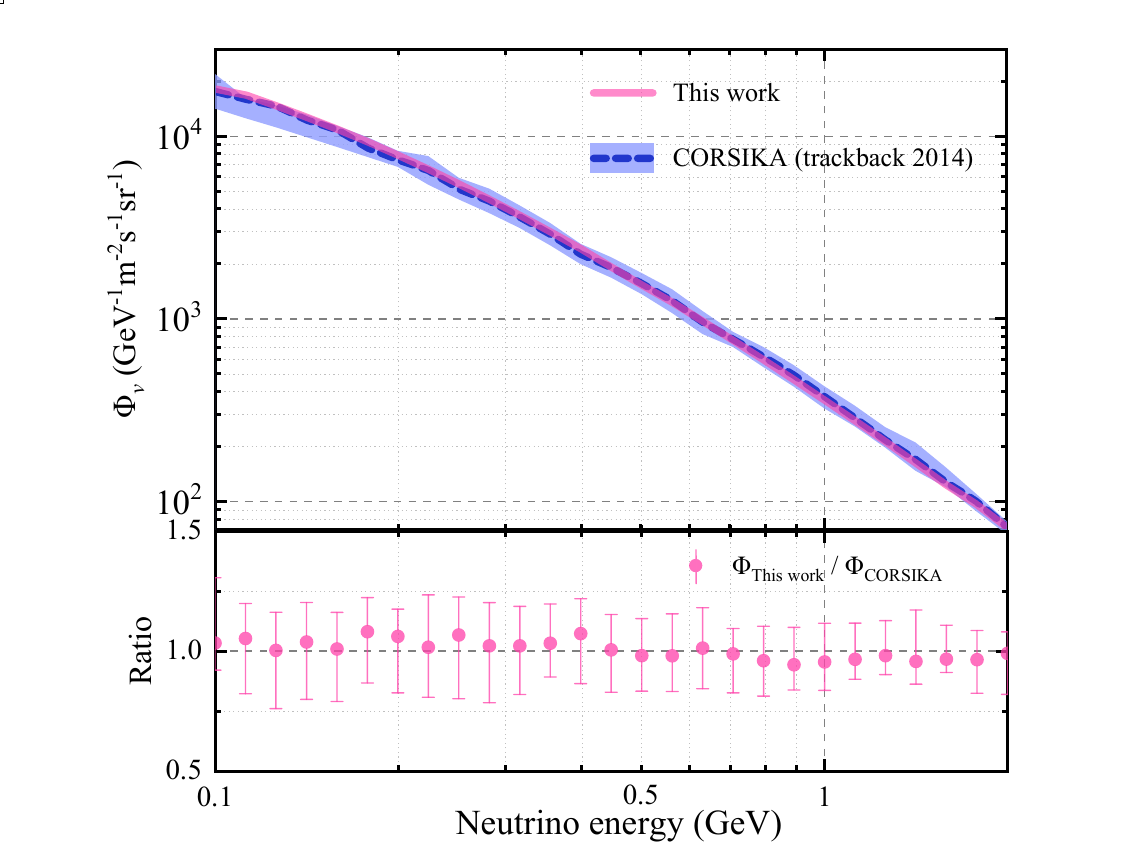}
\caption{Comparison of the total atmospheric neutrino flux at CJPL between this work and the CORSIKA-based result in the energy range 0.1–2 GeV. The lower panel shows the ratio $\Phi_{\mathrm{This\ work}}/\Phi_{\mathrm{CORSIKA}}$.}
\label{fig4}
\end{figure}

The atmospheric neutrino flux estimated for CJPL is in good agreement with the site-specific flux reported by Zhuang et al. \cite{bib:36} based on CORSIKA simulations. Fig.~\ref{fig4} compares the atmospheric neutrino flux calculated in this work with the CORSIKA prediction over the common energy range of 0.1-2 GeV. The ratio, defined as $\Phi_{\mathrm{This\ work}}/\Phi_{\mathrm{CORSIKA}}$, agrees with unity to within 5\% throughout this energy range. An additional contribution to the atmospheric neutrino flux may arise from cosmic-ray muons that stop in the surrounding rock and subsequently decay. Based on the estimate of Wan-Lei Guo \cite{bib:37} and its extrapolation to CJPL conditions, this contribution is expected to be less than 10\% of the atmospheric neutrino flux and is therefore neglected in the present analysis.

\begin{figure}[!htb]
\includegraphics
  [width=0.8\hsize]
  {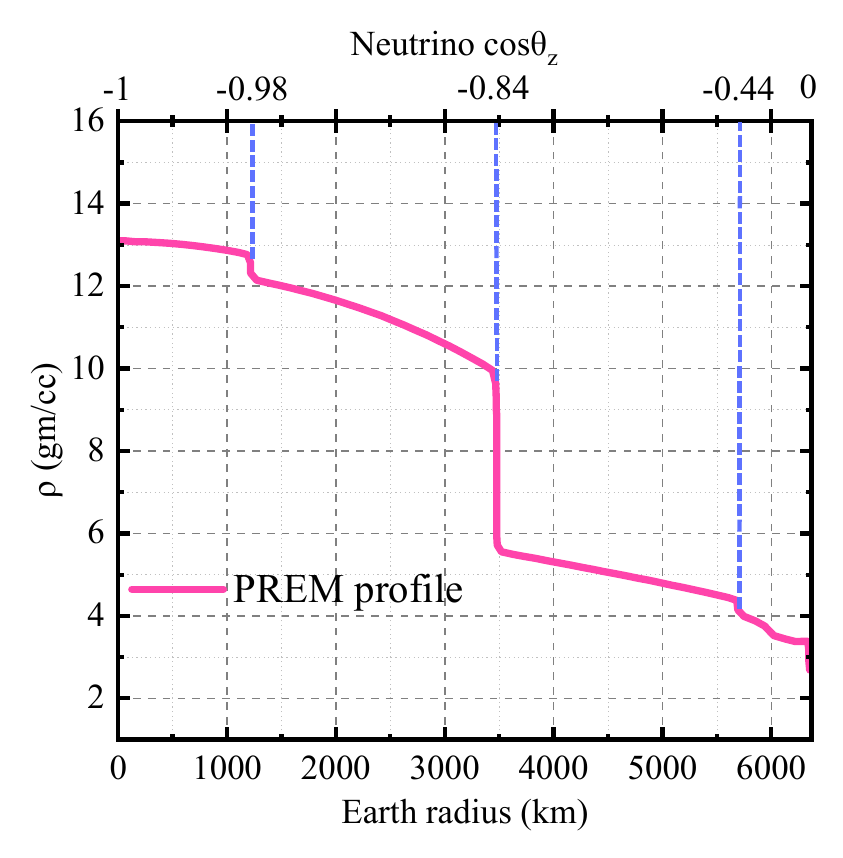}
\caption{The PREM layered density model.}
\label{fig5}
\end{figure}
\begin{figure*}[!htb]
\includegraphics
  [width=0.8\hsize]
  {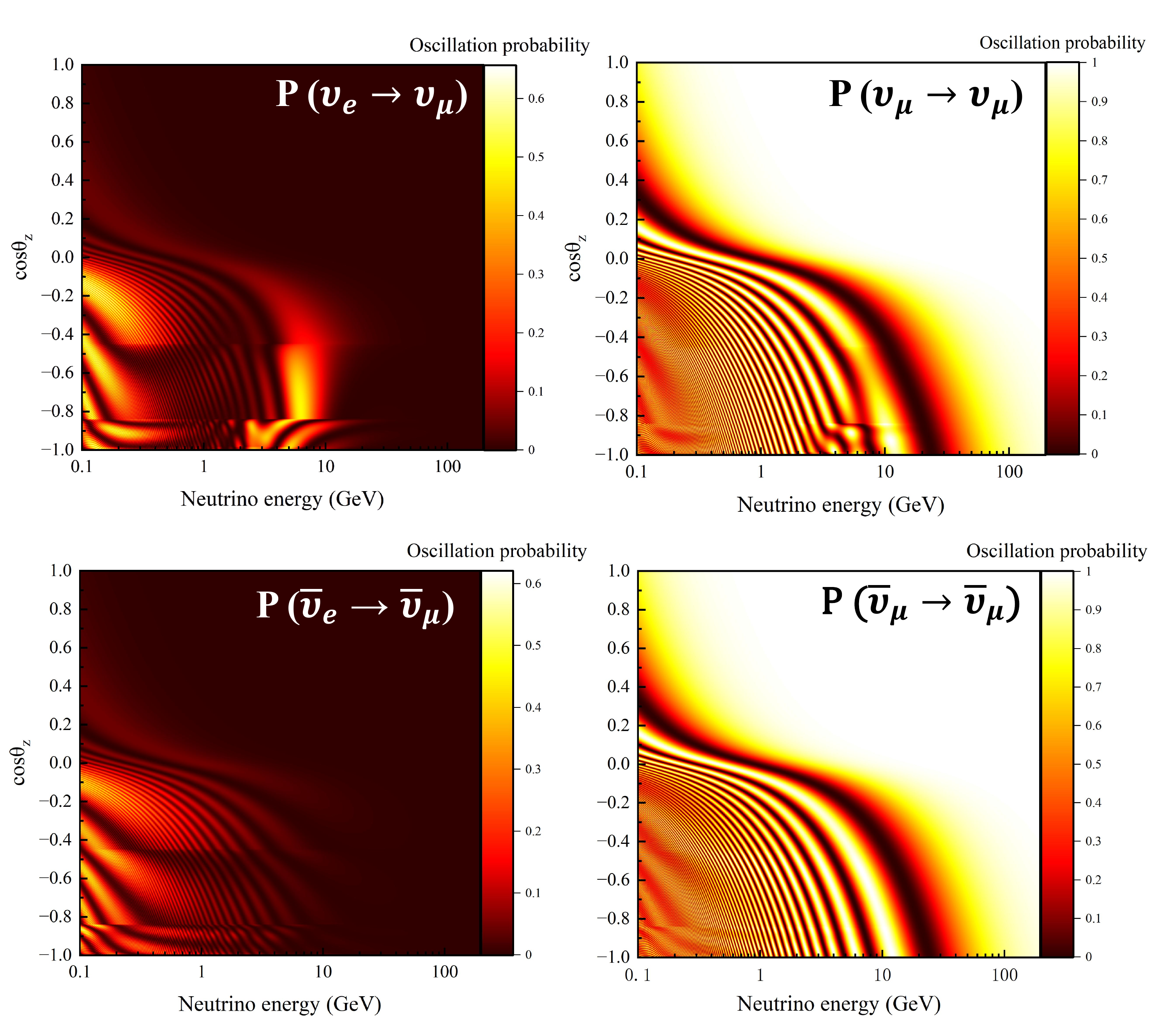}
\caption{Oscillation probabilities for neutrinos (top) and antineutrinos (bottom) after propagation through the Earth to a detector depth of 2.4 km (normal mass ordering). The oscillation parameters are sin$^{2}$$\theta$$_{23}$ = 0.5, sin$^{2}$$\theta$$_{12}$ = 0.304, sin$^{2}$$\theta$$_{13}$ = 0.022, $\Delta$m$^{2}$$_{21}$ = 7.5×10$^{-5}$ eV$^{2}$, $\Delta$m$^{2}$$_{32}$ = 2.4×10$^{-3}$ eV$^{2}$, and $\delta$$_{CP}$ = 0.}
\label{fig6}
\end{figure*}
\section{Neutrino oscillations}\label{sec:3}
Neutrinos have tiny but nonzero masses, and their flavor eigenstates are not identical to their mass eigenstates. A neutrino produced in a definite flavor state is therefore a coherent superposition of mass eigenstates. During propagation, the different mass eigenstates accumulate different phases, and the resulting phase differences change the flavor content of the neutrino state. This phase evolution gives rise to neutrino flavor oscillations over macroscopic distances. In vacuum, the muon-neutrino survival probability can be approximated as
\begin{equation}
P_{\mu\mu} = 1 - \sin^2 2\theta_{23} \sin^2\left(1.27\Delta m_{32}^2\frac {L}{E}\right).
\end{equation}
The oscillation probability therefore depends on the ratio of the neutrino path length $L$ (km) to its energy $E$ (GeV).

When neutrinos propagate through matter, the oscillation probabilities are modified by the Mikheyev--Smirnov--Wolfenstein (MSW) effect. For atmospheric neutrinos traversing the Earth, this matter effect depends on the propagation path and the Earth density profile, and therefore on the zenith-angle. In the present work, three-flavor oscillation probabilities in matter are evaluated using the formalism described in Refs.~\cite{bib:38,bib:39}:
\begin{equation}
P_{\nu_\alpha \rightarrow \nu_\beta}(E, h, \cos\theta_{\mathrm{zenith}})
= \left|
\left(
\mathbf{U}
\prod_{l=1}^{N} \mathbf{X}(L_l, \rho_l, E)
\mathbf{U}^{\dagger}
\right)_{\alpha\beta}
\right|^2.
\end{equation}

\begin{equation}
\mathbf{X} = \sum_{k}
\left[
\prod_{j \neq k} \frac{2E \mathbf{H}_{\mathrm{matter}} - M_j^2 \mathbf{I}}{M_k^2 - M_j^2}
\right]
\exp\left(-i \frac{M_k^2 L}{2E}\right).
\end{equation}

\begin{equation}
\mathbf{H}_{\mathrm{matter}} =
\begin{pmatrix}
\frac{m_1^2}{2E} & 0 & 0 \\
0 & \frac{m_2^2}{2E} & 0 \\
0 & 0 & \frac{m_3^2}{2E}
\end{pmatrix}
+ \mathbf{U}^{\dagger}
\begin{pmatrix}
a & 0 & 0 \\
0 & 0 & 0 \\
0 & 0 & 0
\end{pmatrix}
\mathbf{U}.
\end{equation}

This expression shows that the oscillation probability depends on the neutrino energy, the propagation baseline, and the matter density along the trajectory. For the baseline calculation, we adopt the standard spherical-Earth approximation commonly used in atmospheric neutrino oscillation analyses. Under this approximation, the neutrino path length is determined by the zenith-angle and can be written as \cite{bib:10}:
\begin{equation}
L = \sqrt{(R_0 + L_0)^2 - (R \sin\theta_z)^2} - R \cos\theta_z.
\end{equation}
In this expression, $R_{0}$ denotes the Earth’s radius ($R_{0}=6371,\mathrm{km}$), $L_{0}$ is the typical production altitude of atmospheric neutrinos ($\approx$ $15,\mathrm{km}$). The quantity $R$ is given by $R=R_{0}-d$, where $d$ represents the effective detector depth below the local surface. For CJPL, which is located beneath a high-altitude mountain region, we take $d\approx 2.4~\mathrm{km}$ as the local rock overburden in this geometric approximation. The quantity $\cos\theta_{z}$ denotes the cosine of the atmospheric neutrino zenith-angle.

 To model the matter density along the neutrino trajectory, we use the Preliminary Reference Earth Model (PREM) \cite{bib:41}. Previous studies by Super-K have shown that, for the present purpose, a more finely segmented density profile has only a negligible effect on the oscillation results. We therefore adopt a four-layer Earth model consisting of the inner core, outer core, mantle, and crust. The radii of the layer boundaries, the corresponding densities, and the zenith-angle ranges for trajectories crossing each layer are shown in Fig.~\ref{fig5}.

Fig.~\ref{fig6} shows the oscillation probabilities for muon neutrinos, electron neutrinos, and their antiparticles for the normal mass ordering after propagation through the Earth under CJPL conditions, assuming a detector depth of 2.4 km. The oscillation parameters are sin$^{2}$$\theta$$_{23}$ = 0.5, sin$^{2}$$\theta$$_{12}$ = 0.304, sin$^{2}$$\theta$$_{13}$ = 0.022, $\Delta$m$^{2}$$_{21}$ = 7.5×10$^{-5}$ eV$^{2}$, $\Delta$m$^{2}$$_{32}$ = 2.4×10$^{-3}$ eV$^{2}$, and $\delta$$_{CP}$ = 0. The probability maps exhibit structures near cos$\theta_{z}$ $\approx$ -0.84 and -0.44, where the matter density changes as the neutrino trajectory crosses different Earth layers. For upward-going neutrinos at low energies, matter effects modify the oscillation pattern. These features become weak at high energies ($\gtrsim$ 50 GeV) and for trajectories with cos$\theta_{z}$ close to 1. For muon neutrinos, the oscillation amplitude is enhanced below 1 GeV in regions around cos$\theta_{z}$ $\approx$ -0.2, -0.4, and -0.8, and also in the 2–10 GeV range. For antineutrinos, the trend is reversed, with oscillations suppressed in the corresponding regions.

Furthermore, to illustrate more clearly how the oscillation probabilities for neutrinos and antineutrinos depend on sin$^{2}$$\theta$$_{23}$ and $\Delta$m$^{2}$$_{32}$, we introduce a simplified expression for the oscillation probability in matter \cite{bib:42}:
\begin{equation}
\begin{aligned}
P_{\mu\mu}^{(m)} &\approx 1 - \sin^4 \theta_{23} \sin^2 2\theta_{13}^m \sin^2 \Delta_{31}^m \\
&\quad - \sin^2 2\theta_{23} \left[ \sin^2 \theta_{13}^m \sin^2 \Delta_{21}^m + \cos^2 \theta_{13}^m \sin^2 \Delta_{32}^m \right], \\
P_{e\mu}^{(m)}   &\approx \sin^2 \theta_{23} \sin^2 2\theta_{13}^m \sin^2 \Delta_{31}^m,
\end{aligned}
\label{eq:oscillation_matter}
\end{equation}
where,
\begin{equation}
\begin{aligned}
\Delta_{21}^{m} &= 1.27\delta_{32}\frac{L}{E} \frac{1}{2}
\left[ \frac{\sin 2\theta_{13}}{\sin 2\theta_{13,m}} - 1 - \frac{A}{\delta_{32}} \right], \\
\Delta_{32}^{m} &= 1.27\delta_{32}\frac{L}{E} \frac{1}{2}
\left[ \frac{\sin 2\theta_{13}}{\sin 2\theta_{13,m}} + 1 + \frac{A}{\delta_{32}} \right], \\
\Delta_{31}^{m} &= 1.27\delta_{32}\frac{L}{E}
\left[ \frac{\sin 2\theta_{13}}{\sin 2\theta_{13,m}} \right].
\end{aligned}
\label{eq:delta_m_phases}
\end{equation}
The parameter $\Delta$m$^{2}_{32}$ primarily determines the locations of the oscillation extrema, whereas sin$^{2}\theta_{23}$ mainly controls the oscillation amplitude. The matter potential is $A$ = 2$\sqrt{2}$$G_F$$n_e$$E$. For neutrinos and antineutrinos, the matter term enters with opposite signs, and its impact therefore depends on the mass ordering. As a result, matter effects enhance the oscillation pattern of neutrinos in the normal ordering and that of antineutrinos in the inverted ordering. the exceptionally low CMBg makes it possible to include candidate muon-neutrino events from near-horizontal directions, and even from a limited region with cos$\theta_{z}$ slightly above 0. As indicated in Fig.~\ref{fig6}, for the benchmark parameters the oscillation effect is relatively weak in the region with cos$\theta_{z}$ $\gtrsim$ 0, whereas variations in $\Delta$m$^{2}_{32}$ produce comparatively larger changes there. This feature suggests that extending the usable zenith-angle range toward near-horizontal directions can improve the sensitivity to $\Delta$m$^{2}_{32}$, which is a practical advantage of CJPL for atmospheric neutrino oscillation studies.

\begin{figure}[!htb]
\includegraphics
  [width=0.95\hsize]
  {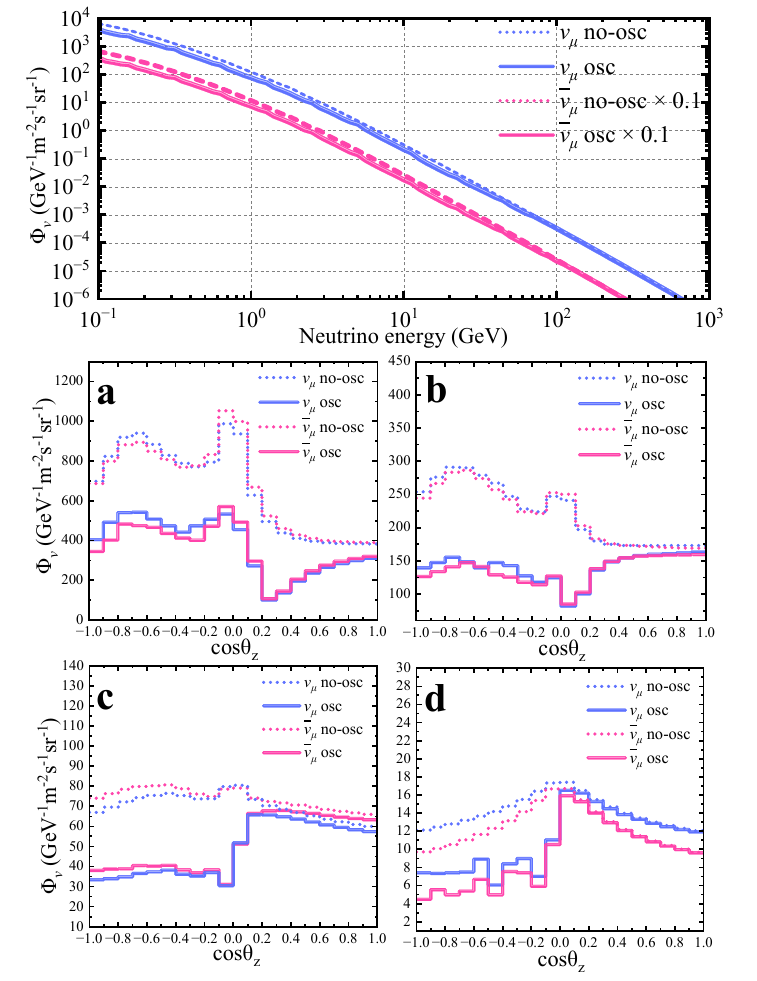}
\caption{Dependence of the CJPL muon neutrino and antimuon neutrino fluxes on energy (top) and zenith-angle (bottom), with and without oscillations. The dashed curves show the unoscillated CJPL fluxes, and the solid curves show the oscillated CJPL fluxes. For visibility, the antimuon neutrino flux in the top panel is scaled by a factor of 1/10. In the bottom panel, $a$, $b$, $c$, and $d$ correspond to the energy ranges 0.1–0.32 GeV, 0.32–1 GeV, 1–3.2 GeV, and $\gtrsim$ 3.2 GeV, respectively. The oscillation parameters are sin$^{2}$$\theta$$_{23}$ = 0.5, sin$^{2}$$\theta$$_{12}$ = 0.304, sin$^{2}$$\theta$$_{13}$ = 0.022, $\Delta$m$^{2}$$_{21}$ = 7.5×10$^{-5}$ eV$^{2}$, $\Delta$m$^{2}$$_{32}$ = 2.4×10$^{-3}$ eV$^{2}$, and $\delta$$_{CP}$ = 0.}
\label{fig7}
\end{figure}

Fig.~\ref{fig7} shows the atmospheric muon-neutrino and antimuon-neutrino fluxes at CJPL as functions of energy and zenith-angle, comparing the cases with and without oscillations for the oscillation parameters sin$^{2}$$\theta$$_{23}$ = 0.5, sin$^{2}$$\theta$$_{12}$ = 0.304, sin$^{2}$$\theta$$_{13}$ = 0.022, $\Delta$m$^{2}$$_{21}$ = 7.5×10$^{-5}$ eV$^{2}$, $\Delta$m$^{2}$$_{32}$ = 2.4×10$^{-3}$ eV$^{2}$, and $\delta$$_{CP}$ = 0. The implementation of neutrino oscillations is described in Section \ref{sec:5}. Since the detectable final-state particles in this analysis are $\mu^{-}$ and $\mu^{+}$, we focus on the spectral shapes of the muon neutrino and antimuon neutrino fluxes. The total muon neutrino flux integrated over the analysis phase space is 1016 m$^{-2}$ s$^{-1}$ sr$^{-1}$ before oscillations and 608 m$^{-2}$ s$^{-1}$ sr$^{-1}$ after oscillations. The corresponding antimuon neutrino flux is 1012 m$^{-2}$ s$^{-1}$ sr$^{-1}$ before oscillations and 576 m$^{-2}$ s$^{-1}$ sr$^{-1}$ after oscillations. As shown in the top panel of Fig.~\ref{fig7}, the oscillation-induced reduction of the muon neutrino flux is concentrated mainly below 50 GeV. The bottom panel shows that the oscillation effect remains significant near the horizontal direction over a broad energy range. For trajectories with cos$\theta_{z}$ close to 1, the oscillation effect is generally weak because of the shorter baseline, although a visible modification remains at a few hundred MeV. The strongest deviations appear around cos$\theta_{z}$ $\approx$ 0.2 for 0.1–0.32 GeV, cos$\theta_{z}$ $\approx$ 0 for 0.32–1 GeV, cos$\theta_{z}$ $\approx$ - 0.1 for 1–3.2 GeV, and cos$\theta_{z}$ $\approx$ -0.2 for E $\gtrsim$ 3.2 GeV. This pattern is governed by the $\Delta$m$^{2}_{32}$-dependent oscillation phase, which shifts the locations of the oscillation minima across different zenith-angle intervals and energy ranges.

\section{Muon event rate and rock-muon flux}\label{sec:4}
\subsection{NuWro simulation of neutrino–matter interactions}

To model muon and pion production in neutrino–matter interactions, the neutrino interaction stage is simulated using the NuWro event generator (v21.09). NuWro provides a detailed description of neutrino–nucleus interactions together with the kinematics of the final-state particles and has been widely used in neutrino interaction studies \cite{bib:43}. In the present study, the generated neutrino sample includes $\nu_{\mu}$, $\bar{\nu}_{\mu}$, $\nu_{e}$, and $\bar{\nu}_{e}$ components, with both charged-current (CC) and neutral-current (NC) interactions considered. For the CC channel, the enabled interaction modes are quasi-elastic scattering (QEL), resonant pion production (RES), deep inelastic scattering (DIS), 
meson-exchange currents (MEC), and coherent scattering (COH). QEL and RES 
dominate for neutrino energies below 3 GeV, whereas DIS becomes the dominant 
interaction mechanism above 3 GeV. For $\nu_{\mu}$ and $\bar{\nu}_{\mu}$ CC interactions, muons are produced directly at the primary interaction vertex. In addition, secondary muons from the decays of $\pi^{+}$ and $\pi^{-}$ produced in neutrino interactions are included. For $\nu_{e}$ and $\bar{\nu}_{e}$ interactions, charged pions produced in the hadronic final states are also allowed to decay and contribute secondary muons. Therefore, the simulated muon sample contains both primary muons from $\nu_{\mu}/\bar{\nu}_{\mu}$ CC interactions and secondary muons from pion decays induced by both muon- and electron-flavor neutrinos.

\begin{figure}[htbp]
\includegraphics
  [width=1\hsize]
  {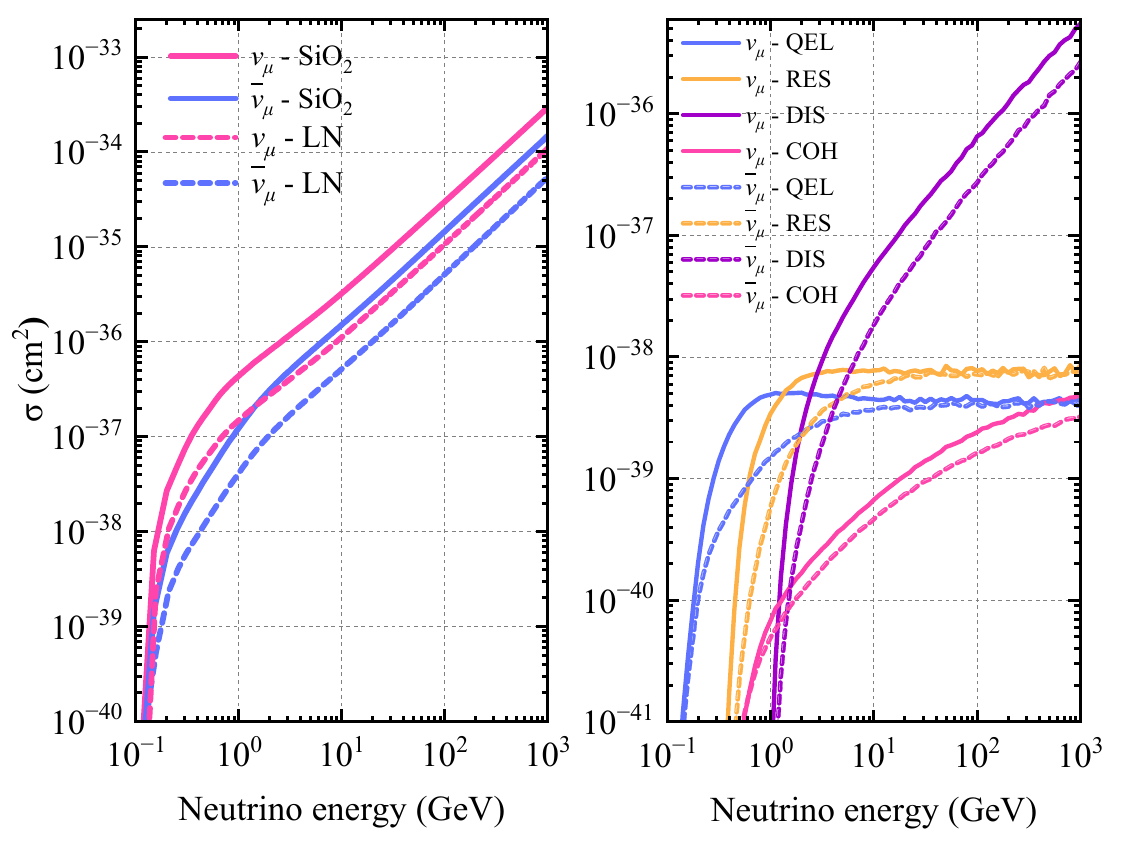}
\caption{NuWro CC cross sections for muon neutrinos and antimuon neutrinos interacting in rock and in liquid nitrogen. The right panel shows the cross sections in rock separated by interaction channel. }
\label{fig8}
\end{figure}

The axial mass M$_{A}$ is set to 1.03 GeV, and neutrino–nucleus interactions are modeled using the relativistic Fermi gas (RFG) model. Fig.~\ref{fig8} shows the total CC cross sections from NuWro for muon neutrinos and antimuon neutrinos with energies above 0.1 GeV interacting in rock (SiO$_{2}$, density = 2.8 g/cm$^{3}$) and with liquid nitrogen (N$_{2}$, density = 0.808 g/cm$^{3}$). The CC cross sections on SiO$_2$ are approximately a factor of two larger than those on liquid nitrogen for both $\nu_{\mu}$ and $\bar{\nu}_{\mu}$. Over the energy range considered here, the $\nu_{\mu}$ cross section is also substantially larger than the corresponding $\bar{\nu}_{\mu}$ cross section.

\begin{figure}[htbp]
\includegraphics
  [width=1\hsize]
  {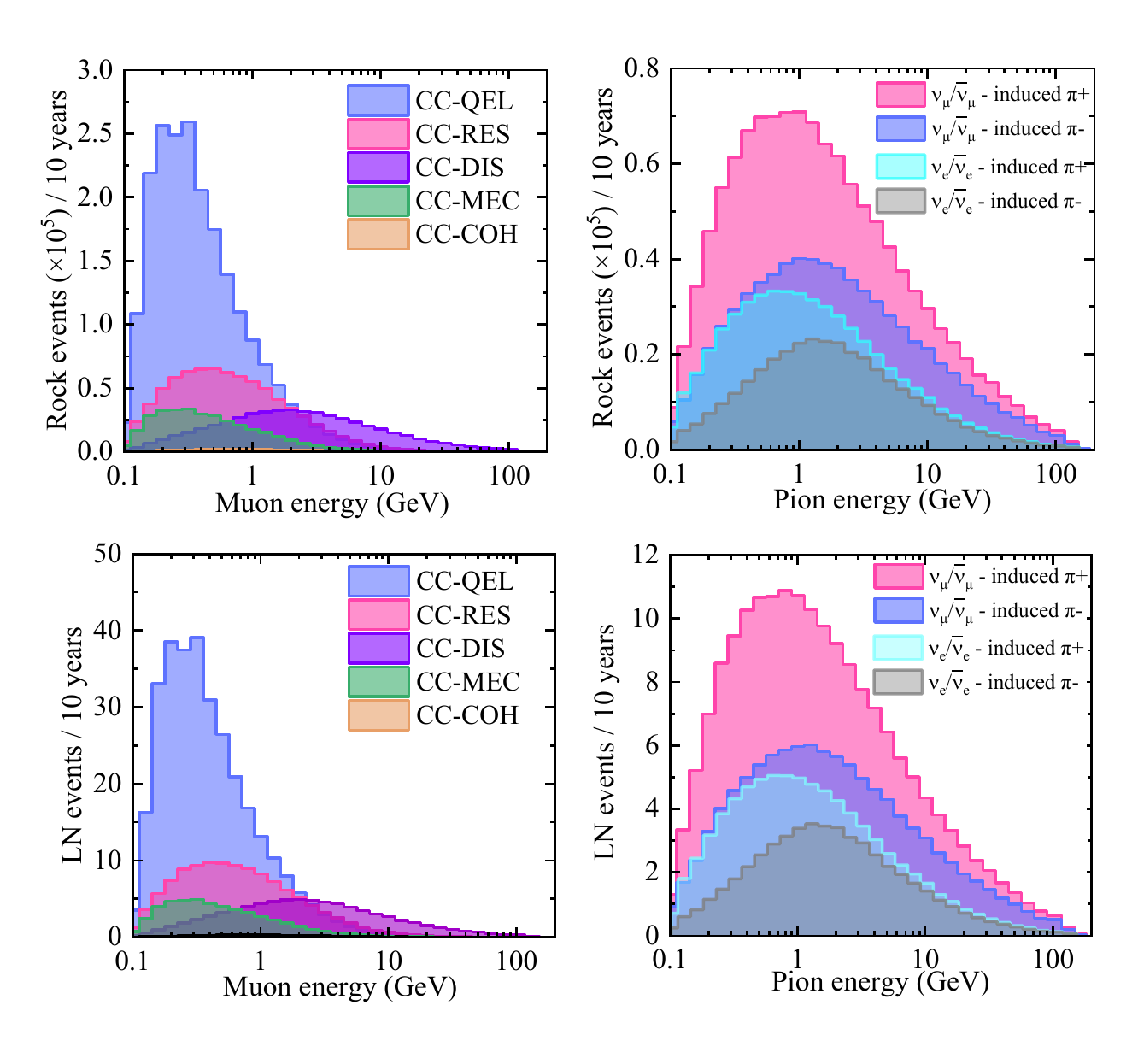}
\caption{Energy spectra of $\mu^{\pm}$ (left) and $\pi^{\pm}$ (right) produced in rock (top) and liquid nitrogen (bottom) at CJPL for a 10-year exposure. The muon spectra are separated by CC and NC interaction channels, while the pion spectra include contributions from both $\nu_{\mu}/\bar{\nu}_{\mu}$ and $\nu_{e}/\bar{\nu}_{e}$ interactions. The ordinates of the upper panels are scaled by $10^{5}$.}
\label{fig9}
\end{figure}

The four panels of Fig.~\ref{fig9} show the energy distributions of $\mu^{\pm}$ and $\pi^{\pm}$ produced in neutrino interactions within rock and liquid nitrogen. The $\mu^{\pm}$ distributions contain both primary muons from $\nu_{\mu}/\bar{\nu}_{\mu}$ CC interactions and secondary muons from $\pi^{\pm}$ decays. The $\pi^{\pm}$ distributions include contributions from both muon- and electron-flavor neutrino interactions, including both CC and NC processes. The muon distributions are separated by interaction channel, while the pion distributions are separated by neutrino flavor and pion charge. The vertical axis is normalized to the expected number of interactions at CJPL for a 10-year exposure using Eq.~\eqref{eq10}. In this equation, $T$ is the exposure time, $4\pi$ represents the integration over solid angle, and $A$ is the number of target nucleons in the corresponding interaction volume, with $A_{rock}$ $\approx$ 8.1×10$^{36}$ for the surrounding rock volume of 200 m × 200 m × 120 m. In Eq.~\eqref{eq10}, $\Phi_{\nu_\alpha}(E)$ and $\Phi_{\bar{\nu}_\alpha}(E)$ denote the oscillated neutrino and antineutrino fluxes for flavor $\alpha$ ($\alpha=e,\mu$), while $\sigma_{\nu_\alpha}(E)$ and $\sigma_{\bar{\nu}_\alpha}(E)$ are the 
corresponding total interaction cross sections.

\begin{equation}\label{eq10}
N = 4\pi \cdot T \cdot A
\int
\sum_{\alpha=e,\mu}
\left[
\Phi_{\nu_\alpha}(E)\sigma_{\nu_\alpha}(E)
+
\Phi_{\bar{\nu}_\alpha}(E)\sigma_{\bar{\nu}_\alpha}(E)
\right]
\, dE .
\end{equation}

Fig.~\ref{fig9} shows that most secondary muons are produced with energies around 500 MeV, with the QEL and RES channels providing the dominant contributions. For electron-flavor neutrinos, CC interactions do not directly produce muons; instead, secondary muons arise from charged pions produced in the hadronic final states and subsequently decaying into muons. The pion spectra further indicate that $\pi^{+}$ and $\pi^{-}$ decays from both CC and NC interactions in the surrounding rock make a non-negligible contribution to the total rock-muon yield. Although the initial muon yield from neutrino interactions inside the liquid-nitrogen volume is much smaller than that from interactions in the surrounding rock, the two components are affected very differently by subsequent transport. Muons produced in rock lose energy through ionization and radiative processes during propagation, and only a small fraction can reach the detector volume. By contrast, muons produced inside the liquid-nitrogen cryostat are generated directly in the active volume and therefore suffer little propagation loss before detection. As a result, the relative importance of the liquid-nitrogen component becomes significantly enhanced at the detector level compared with that at production. The transport stage and the subsequent detector-level selection are modeled with Geant4, as described in the next section.

\subsection{Geant4 simulation of muon and pion transport}

The secondary $\mu^{\pm}$ and $\pi^{\pm}$ generated with NuWro are passed to Geant4 as inputs in order to simulate pion decays and muon transport in the surrounding rock and in the liquid-nitrogen cryostat. We use Geant4 version 4.10.06 Hadronic interactions of pions and other secondary hadrons are modeled within the standard Geant4 hadronic physics framework, while the transport of muons and charged hadrons through rock and liquid nitrogen is treated consistently in the same simulation chain. The decay of unstable secondary particles, including $\pi^{\pm}$, is handled by the Geant4 decay processes.

\begin{figure}[!htb]
\includegraphics
  [width=0.9\hsize]
  {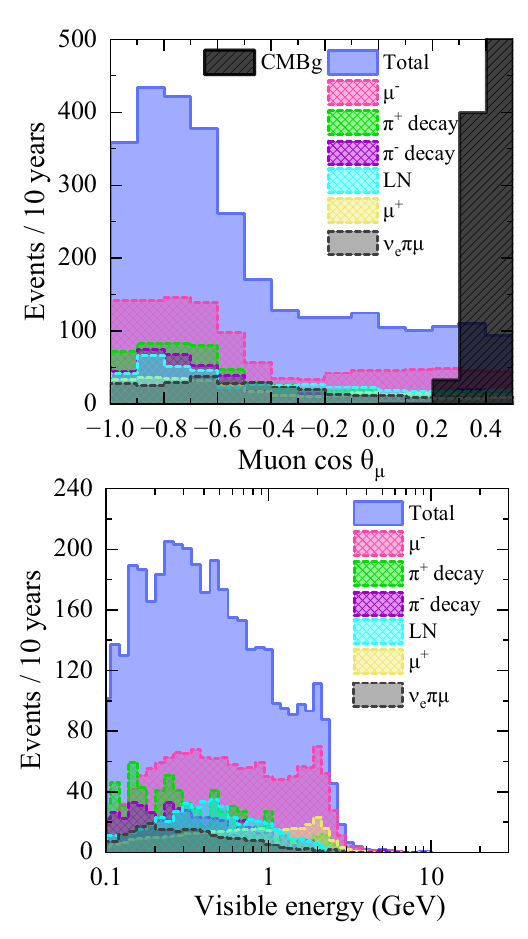}
\caption{Geant4 simulation results for the liquid-nitrogen cryostat, normalized to a 10 year exposure in order to reduce Monte Carlo statistical fluctuations. The top panel shows the zenith-angle distribution of muons in the range $\cos\theta_{\mu}\in[-1,0.3]$. The bottom panel shows the corresponding visible-energy distribution, where the visible energy is represented by the muon energy deposited in the detector volume. Here, $\mu^{\pm}$ denotes muons and antimuons produced directly in CC interactions of neutrinos in the surrounding rock, $\pi^{\pm}$ denotes secondary muons and antimuons originating from the decays of pions produced in $\nu_{\mu}/\bar{\nu}_{\mu}$ interactions, $LN$ denotes the total visible energy of events within the liquid nitrogen volume, $\nu_e \pi \mu$ denotes $\mu^\pm$ originating from $\pi^\pm$ decays induced by $\nu_e$ or $\bar{\nu}_e$ interactions and the black hatched region represents the contribution from the residual CMBg.}
\label{fig10}
\end{figure}

Fig.~\ref{fig10} presents the Geant4 simulation results for the liquid-nitrogen volume, normalized to a 10-year exposure to suppress Monte Carlo statistical fluctuations. The top panel shows the distribution of the muon zenith-angle cosine, $\cos\theta_{\mu}$, at the entrance to the liquid-nitrogen volume, including the extended angular region up to $\cos\theta_{\mu}=0.3$. Since the muon charge is not distinguished in the present analysis, $\mu^{-}$ and $\mu^{+}$ are combined in the final event sample. The projected event sample contains 2731 rock-induced muon events and 493 events originating from neutrino interactions inside the liquid-nitrogen volume over a 10-year exposure. The rock-induced sample is dominated by $\mu^{-}$ produced directly in CC interactions in the surrounding rock, followed by muons from $\pi^{+}$ decays, muons from $\pi^{-}$ decays, and finally $\mu^{+}$ produced directly in $\bar{\nu}_{\mu}$ CC interactions. The contributions from $\pi^{+}$ and $\pi^{-}$ decays exceed that from directly produced $\mu^{+}$ because both  $\nu_{\mu}$ and $\bar{\nu}_{\mu}$ can produce charged pions through RES and DIS processes, while the $\nu_{\mu}$ CC cross section is significantly larger than that of $\bar{\nu}_{\mu}$. The $\nu_e \pi \mu$ events in Fig.~\ref{fig10} refer to $\mu^-/\mu^+$ from pion decays induced by $\nu_e/\bar{\nu}_e$. Since a dedicated particle identification algorithm and effective suppression techniques are not applied in the present analysis, these events are conservatively included as a background component. The number of such background events originating from both the rock and the liquid nitrogen, as estimated from simulation, amounts to 278. In addition, the residual CMBg is confined to the upper edge of the extended angular region, contributing 33 events in the interval $0.2 < \cos\theta_{\mu} < 0.3$ over a 10-year exposure. This estimate is obtained by scaling a limited cosmic-ray muon sample and is therefore dominated by Poisson counting statistics, corresponding to an uncertainty of approximately $\pm13$ events. Consequently, the residual CMBg is localized near the boundary of the extended zenith-angle region rather than being uniformly distributed over the full interval. The bottom panel of Fig.~\ref{fig10} shows the visible-energy spectrum above 0.1 GeV. In the present analysis, the visible energy is defined as the total energy deposited by all detectable final-state particles in the liquid-nitrogen volume, including contributions from both muons and hadronic particles, and is used as the observable proxy for the neutrino event energy. For muons above the Cherenkov threshold, this deposited energy is expected to be correlated with the Cherenkov light yield, although the relation is not strictly one-to-one, especially for low-energy or stopping muons near threshold. 

Based on the Geant4 simulation, we obtain the detection fractions for muons and antimuons originating from the surrounding rock, including both those produced directly in CC interactions and those arising from pion decays. The corresponding fractions are 3.02 × 10$^{-4}$ and 3.45 × 10$^{-4}$, respectively. We also evaluate the effective area of the liquid-nitrogen cryostat for rock muons using the muon-projection method. The corresponding expression is given in Eq.~\eqref{eq11}, where $\theta_{\mu}$ and $\phi$ are the muon zenith-angles and azimuth-angles, and $R$ and $H$ are the radius and height of the liquid-nitrogen vessel, respectively. 
Using this method, we obtain an effective area of 378 m$^{2}$ for the zenith-angle cosine range $\cos\theta_{\mu}\in[-1,0.3]$. The corresponding rock-muon flux is $(3.65 \pm 1.00)\times10^{-13}$ cm$^{-2}$ s$^{-1}$ sr$^{-1}$, and the muon yield in the $1725~\mathrm{m^{3}}$ liquid-nitrogen volume is $(0.13 \pm 0.034)$ day$^{-1}$, both evaluated under the same angular acceptance $\cos\theta_{\mu}\in[-1,0.3]$.

\begin{equation}
\label{eq11}
\begin{split}
S ={}& \iint \Big( |\sin\theta_\mu| \cdot |\cos\phi| \cdot 2RH \\
    & \qquad + |\cos\theta_\mu| \cdot \pi R^2 \Big)\,
    d(\cos\theta_\mu)\, d\phi .
\end{split}
\end{equation}

\begin{figure*}[!htb]
\includegraphics
  [width=0.9\hsize]
  {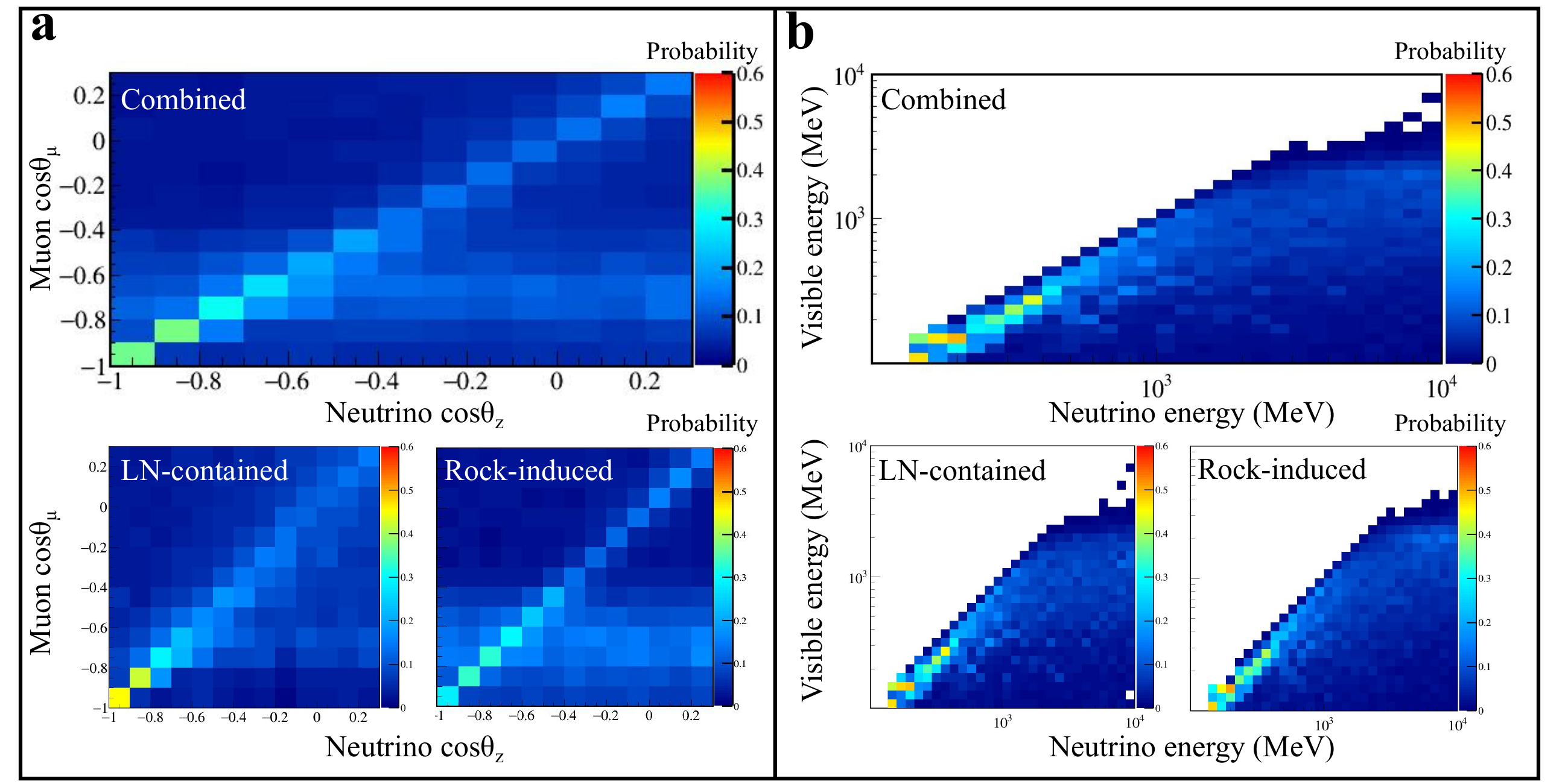}
\caption{Probability migration matrices from true neutrino variables to the observable quantities of detectable muons. (a) Angular correlation matrices, representing the conditional probability ($P(\cos\theta_{\mu}\mid\cos\theta_{z})$), including the combined muon sample, liquid-nitrogen muon events, and rock-induced muon events. (b) Energy correlation matrices, representing the conditional probability ($P(E_{\mathrm{vis}}\mid E_{\nu})$). The color bars on the right indicate the corresponding conditional probabilities.}
\label{fig11}
\end{figure*}
\subsection{Muon event characterization and acceptance}

To further characterize the oscillation analysis sample, we investigate the relationship between the true neutrino energy and zenith-angle cosine of the parent neutrino and the visible energy and zenith-angle cosine of the detectable muon, together with the event acceptance determined by the detector geometry. The selected sample includes primary muons produced in $\nu_{\mu}/\bar{\nu}_{\mu}$ CC interactions and secondary muons originating from charged pion decays induced by $\nu_{\mu}/\bar{\nu}_{\mu}$ and $\nu_{e}/\bar{\nu}_{e}$ interactions via CC and NC processes. Contributions from $\nu_{e}/\bar{\nu}_{e}$ interactions and residual cosmic-ray muons are treated as backgrounds and will be discussed in Sec.~\ref{sec:5}.

\begin{figure}[!htb]
\includegraphics
  [width=0.75\hsize]
  {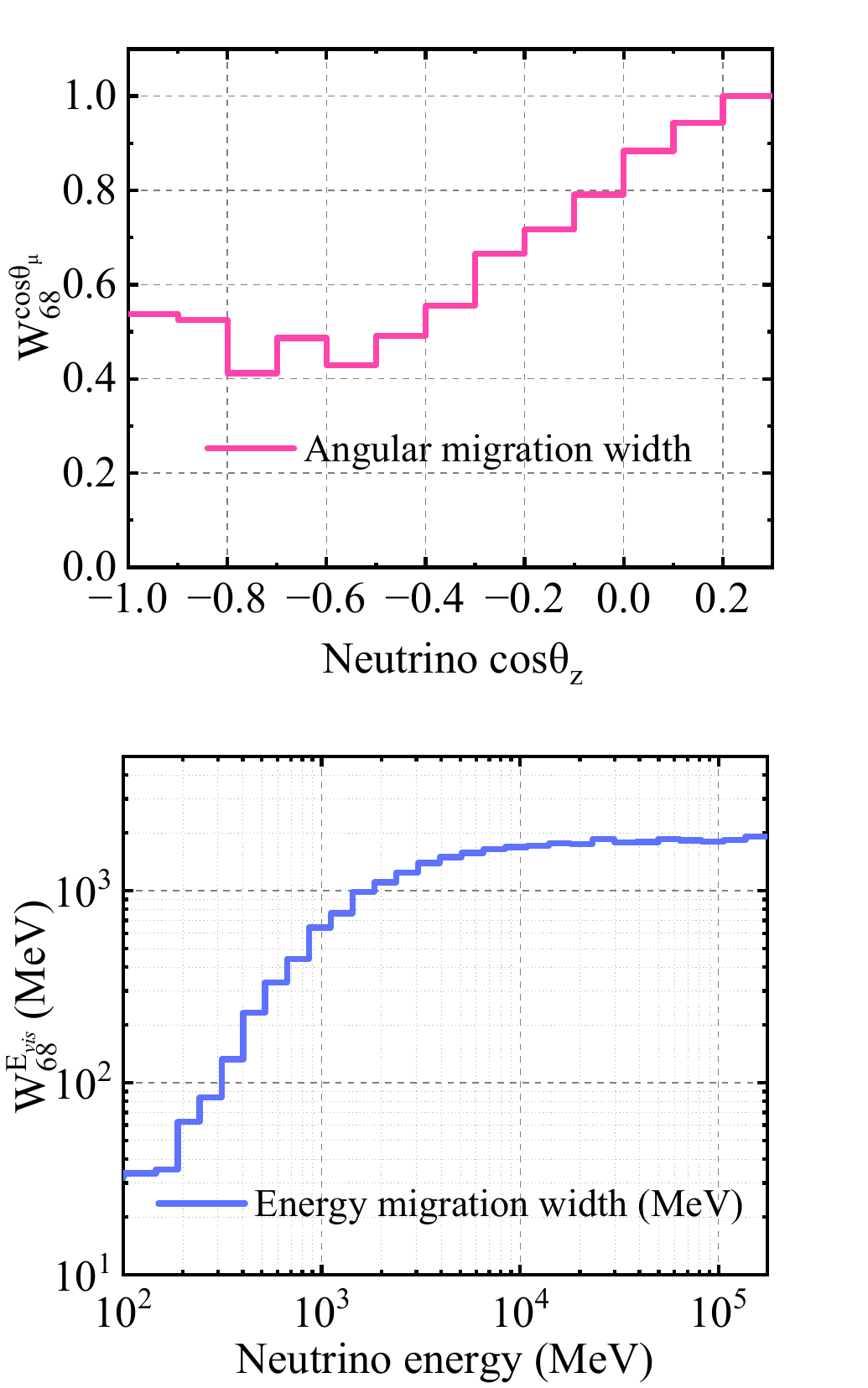}
\caption{Effective angular and energy migration widths derived from the neutrino-to-muon response matrices. The widths are quantified using the 68\% containment interval, where the 16th and 84th percentiles define the lower and upper limits of the migration distributions in each bin.}
\label{fig12}
\end{figure}

Fig.~\ref{fig11} presents the migration matrices relating the true neutrino variables to the reconstructed muon observables. Each matrix gives the conditional probability that a neutrino with a given true energy or zenith angle is reconstructed in a particular muon energy or zenith-angle bin. Fig.~\ref{fig11}(a) shows the migration matrix between the true neutrino zenith-angle cosine and the zenith-angle cosine of the detectable muon. The angular migration probability is defined as: \[
P(\cos\theta_{\mu}\mid\cos\theta_{z})
=
\frac{N(\cos\theta_{z},\,\cos\theta_{\mu})}
{\sum_{\cos\theta_{\mu}} N(\cos\theta_{z},\,\cos\theta_{\mu})}
\]. The upper rectangular panel shows the combined event sample, including both 
liquid-nitrogen-contained and rock-induced events, while the two lower square 
panels present the two components separately. The accumulation band around $\cos\theta_\mu\approx-0.7$ originates from the combined contributions of neutrino-muon angular responses at different energy ranges. Low-energy neutrinos (below 1 GeV) produce muons with large kinematic angular deviations, resulting in weak correlations between the muon and incident neutrino directions. After the angular migration is expressed in terms of $\cos\theta_\mu$, and owing to the nonlinear relationship between the angle and its cosine, these low-energy contributions form a broad distribution mainly extending over the large zenith-angle region ($\cos\theta_\mu\lesssim-0.4$), appearing as an accumulation band around $\cos\theta_\mu\approx-0.7$. In contrast, higher-energy neutrinos (above 1 GeV) preserve stronger directional correlations, producing the diagonal structure where $\cos\theta_\mu$ follows $\cos\theta_\nu$. Therefore, the observed band structure is determined by the coexistence of the broad low-energy angular migration component and the direction-correlated high-energy component. The smearing is more pronounced for rock-induced events because the muons undergo additional propagation through the surrounding rock before entering the detector. Fig.~\ref{fig11}(b) shows the corresponding correlation matrix between the true neutrino energy and the visible energy of the muon. The energy migration probability is defined as: \[
P(E_{\mathrm{vis}}\mid E_{\nu})
=
\frac{N(E_{\nu},\,E_{\mathrm{vis}})}
{\sum_{E_{\mathrm{vis}}} N(E_{\nu},\,E_{\mathrm{vis}})}
\]. For low-energy events below about 1000 MeV, the true neutrino energy remains strongly correlated with the muon visible energy, while this correlation weakens at higher energies. Fig.~\ref{fig11} also shows that events produced inside the liquid-nitrogen volume remain closer to the original interaction kinematics, whereas rock-induced events exhibit more pronounced smearing and broadening due to the additional propagation in rock before reaching the detector.

To further quantify the corresponding migration effects, the effective angular 
and energy widths are evaluated from the conditional distributions and shown 
in Fig.~\ref{fig12}. Since the migration distributions are generally non-Gaussian, the 68\% containment width is used \cite{bib:44}:
\begin{equation}
W_{68}=q_{84}-q_{16},
\end{equation}
where $q_{16}$ and $q_{84}$ denote the 16th and 84th percentiles of the 
corresponding migration distributions. For the angular migration width, $W_{68}^{\cos\theta_\mu}$ represents the 68\% containment width of the muon zenith-angle cosine distribution for a given incident neutrino zenith-angle cosine, reflecting the kinematic deviation between the neutrino and the produced muon. As shown in Fig.~\ref{fig12}, the angular migration width remains relatively stable in the region of $\cos\theta_\nu<-0.4$ and increases toward larger $\cos\theta_\nu$. This behavior results from the energy-dependent neutrino-muon angular response discussed above. In the region of $\cos\theta_\nu\approx-0.7$, the low- and high-energy components contribute to similar muon angular regions, leading to a relatively concentrated $\cos\theta_\mu$ distribution and a smaller migration width. For other neutrino directions, the coexistence of the broad low-energy migration component and the direction-correlated high-energy component leads to an increased spread of the muon angular distribution. For the energy migration width, $W_{68}^{E_{vis}}$ represents the 68\% containment width of the visible-energy distribution of detectable muons for a given incident neutrino energy. The absolute energy width increases rapidly at low and intermediate energies and gradually approaches a plateau at high energies. This behavior is related to the increasing spread of muon energies and energy losses for higher-energy neutrino interactions. At sufficiently high energies, the visible-energy response becomes limited by the muon energy-loss processes in surrounding rock and the finite energy-deposition range in the liquid-nitrogen detector.

Because of the cylindrical detector geometry, a small fraction of muon events either traverse only a very short distance inside the detector or are produced too close to the detector wall. We therefore study the acceptance of the liquid-nitrogen cryostat for muon events as a function of incident angle and visible energy. This acceptance is treated here as a geometry-based measure of event usability.

\begin{figure}[htbp]
\includegraphics
  [width=1\hsize]
  {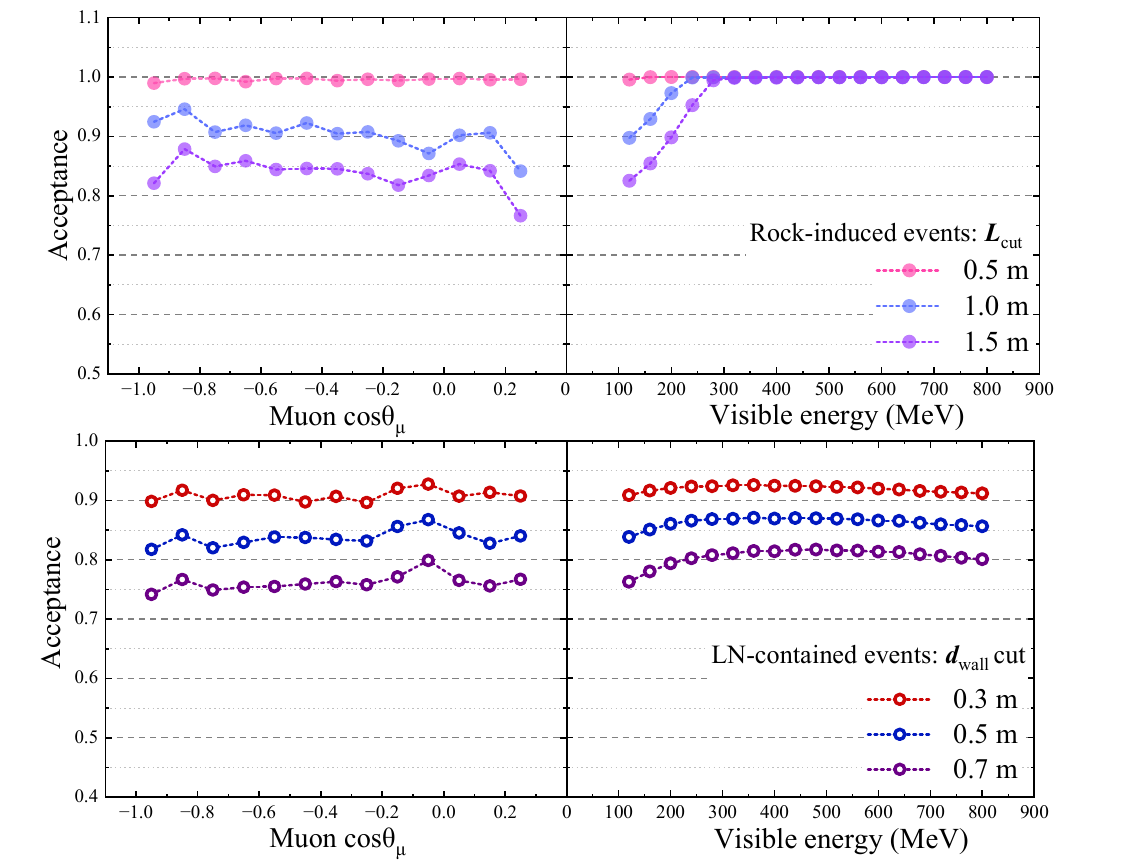}
\caption{Dependence of detector acceptance on geometry-based selection cuts for muon events. The upper panels show rock-induced muon events, for which the track length in the liquid-nitrogen volume, $L_{\mathrm{track}}$, is used as the selection variable. The lower panels show liquid-nitrogen-contained muon events, for which the distance from the event vertex to the inner detector wall, $d_{\mathrm{wall}}$, is used as the selection variable. The left panels present the acceptance as a function of muon zenith-angle cosine, and the right panels show the acceptance as a function of visible energy.}
\label{fig13}
\end{figure}

For external muon events originating from the surrounding rock, the relevant geometric quantity is the muon path length inside the liquid-nitrogen volume. We therefore use the track length in liquid nitrogen, $L_{\mathrm{track}}$, as the selection variable and reject events with $L_{\mathrm{track}} < L_{\mathrm{cut}}$. The upper panels of Fig.~\ref{fig13} show the acceptance of rock-induced muon events as a function of muon zenith-angle cosine and visible energy for several choices of $L_{\mathrm{cut}}$. Here, the angular acceptance is defined as the fraction of events passing the cut in a given $\cos\theta_{\mu}$ bin, while the visible-energy acceptance is the corresponding fraction in a given visible-energy bin. A cut of $L_{\mathrm{cut}} = 0.5~\mathrm{m}$ is too loose to provide effective event selection. As $L_{\mathrm{cut}}$ increases, short-track events crossing the detector edge or corners are progressively removed, along with part of the low-energy sample, and the event statistics decrease accordingly. In the following analysis, we adopt $L_{\mathrm{cut}} = 1~\mathrm{m}$ for rock-induced events. With this choice, the cut mainly affects the lowest visible-energy region; for events below $200~\mathrm{MeV}$, the suppression remains modest, and the acceptance stays at roughly the 90\% level over the full zenith-angle interval. After applying this cut, 2485 rock-induced muon events remain for a 10-year exposure.

For events produced inside the liquid-nitrogen volume, the relevant geometric quantity is no longer the track length, but the distance from the interaction vertex to the detector wall. We therefore use the minimum distance from the event vertex to the inner detector wall, $d_{\mathrm{wall}}$, as the selection variable. The lower panels of Fig.~\ref{fig13} show the acceptance of liquid-nitrogen-contained muon events as a function of angle and visible energy for different choices of $d_{\mathrm{wall}}$. Since neutrino interactions in liquid nitrogen are approximately uniform throughout the detector volume, increasing $d_{\mathrm{wall}}$ leads to a gradual overall reduction in acceptance, rather than a strong distortion in a particular angular or energy region. In the following analysis, we adopt $d_{\mathrm{wall}} = 0.3~\mathrm{m}$ for liquid-nitrogen-contained events, which retains 91.1\% of the events. After applying this cut, 449 liquid-nitrogen-contained events remain for a 10-year exposure.

The acceptance study shown in Fig.~\ref{fig13} is included in the final definition of the oscillation analysis sample. All Geant4 event mapping, observable distributions, and sensitivity calculations are based on the sample after these cuts are applied. The acceptance is not fully uniform over the analysis space. For rock-induced events, the track-length cut mainly suppresses the lowest visible-energy region and the edge bins of the extended zenith-angle range, whereas for liquid-nitrogen-contained events the $d_{\mathrm{wall}}$ cut produces a smoother reduction, closer to a fiducial-volume effect. Overall, these cuts mainly lead to modest shape changes and some loss of statistics, rather than a strong suppression of the full analysis space.

\section{Oscillation parameter analysis}\label{sec:5}
\subsection{Analysis methods}
\begin{figure}[!htb]
\includegraphics
  [width=1\hsize]
  {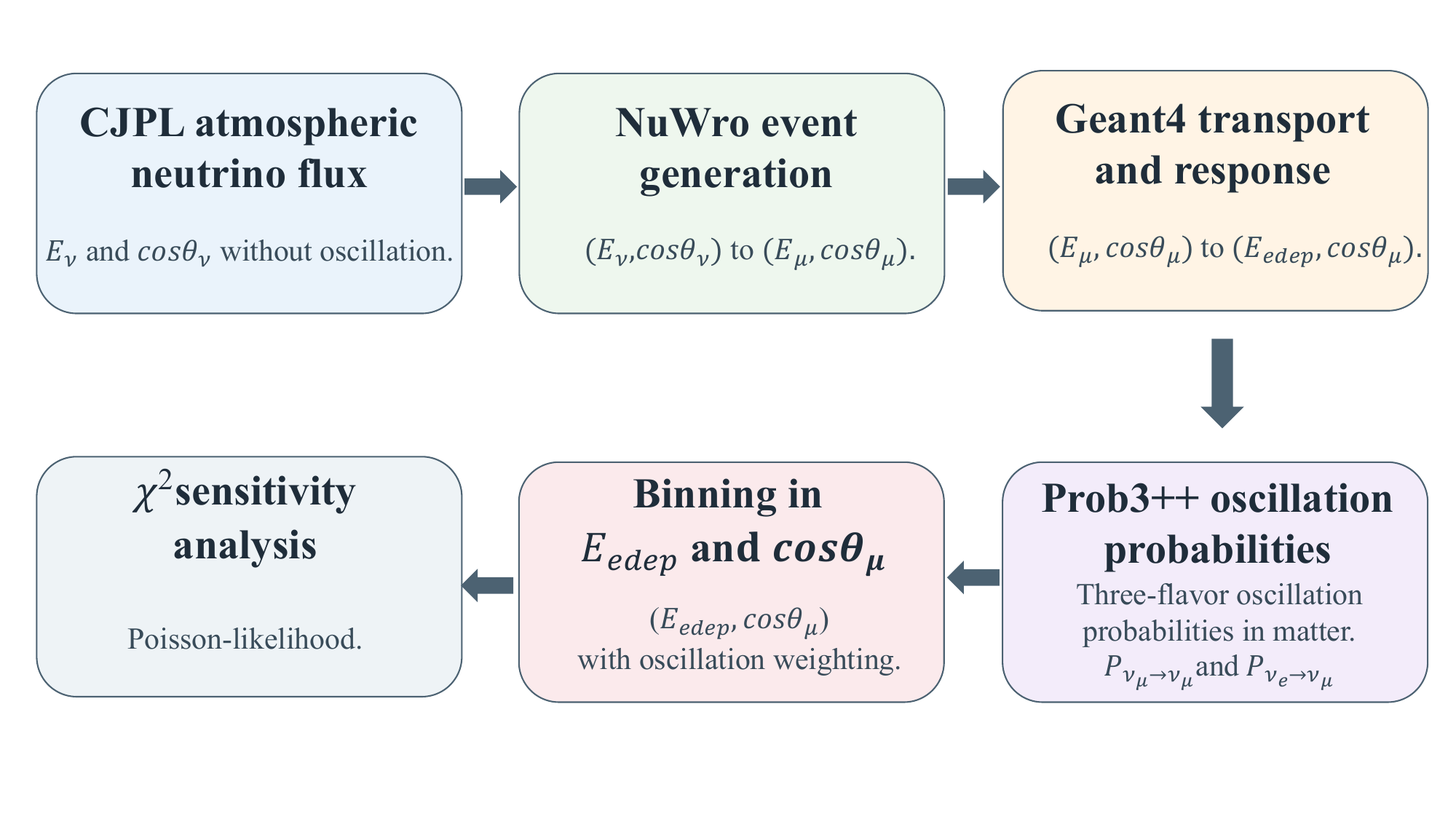}
\caption{Workflow of the oscillation parameter analysis. }
\label{fig14}
\end{figure}

In this study, we evaluate the sensitivity of the CJPL liquid-nitrogen cryostat to the oscillation parameters using a simulation chain based on Prob3++, NuWro, and Geant4. Using NuWro together with the atmospheric neutrino flux derived in Section~\ref{sec:2}, we generate an unoscillated atmospheric neutrino sample corresponding to 100 years of exposure in a 200 m × 200 m × 120 m rock volume surrounding the detector. For each interaction, we record the incident neutrino energy and interaction vertex, together with the energies, directions, and production positions of the secondary muons and pions. These secondary particles are then propagated event by event with Geant4, which simulates pion decays and muon transport and records the muons entering the liquid-nitrogen vessel, including their visible energies and zenith-angles. A visible-energy threshold of 0.1 GeV is applied, and the usable zenith-angle range is restricted to cos$\theta_{\mu}$ $\in$ [-1, 0.3]. The workflow of the oscillation parameter analysis is shown in Fig.~\ref{fig14}.

The Geant4 input and output records are then matched to the corresponding NuWro neutrino inputs and secondary-particle outputs. In this way, under the no-oscillation assumption, we construct an event-by-event response mapping between the incident neutrino truth variables and the muon observables in the liquid-nitrogen vessel. Prob3++ is then used to calculate the three-flavor oscillation probabilities as functions of neutrino flavor, energy, and zenith-angle. For each generated event, an oscillation weight is assigned according to the parent neutrino flavor and its truth kinematics, and the weighted contribution is accumulated in the observable space of visible energy and muon zenith-angle cosine according to

\begin{equation}
\begin{gathered}
N_{ij}^{\text{obs}} = N_{ij}^{\nu_{\mu}} + B_{ij}^{\nu_e} + \delta_i^{\mathrm{CR}}B_{i}^{\rm CR}, \\
N_{ij}^{\nu_{\mu}} = \sum_k \sum_{\alpha=e,\mu} P_{\alpha\mu}(E_{\nu,k}, \cos\theta_{z,k}) \, I_{ij}^{\nu_{\mu}}(E_{\text{vis},k}, \cos\theta_{\mu,k}), \\
B_{ij}^{\nu_e} = \sum_k \sum_{\alpha=e,\mu} P_{\alpha e}(E_{\nu,k}, \cos\theta_{z,k}) \, I_{ij}^{\nu_e}(E_{\text{vis},k}, \cos\theta_{\mu,k}).
\end{gathered}
\end{equation}

Here, $N_{ij}^{\nu_{\mu}}$ denotes the $\nu_\mu$-like contribution, including primary muons from $\nu_\mu$ CC interactions and secondary muons from charged-pion decays following both CC and NC interactions. $B_{ij}^{\nu_e}$ denotes the background from muons arising from pion decays in $\nu_e/\bar{\nu}_e$ interactions, and $B_{i}^{\rm CR}$ denotes the residual cosmic-ray muon background. The binary selection factor $\delta_i^{\mathrm{CR}}$ equals 1 for zenith-angle bins that contain residual cosmic-ray events and 0 otherwise, indicating the bins where the cosmic-ray background component is included. In the present analysis, $\delta_i^{\mathrm{CR}} = 1$ only for $i = 13$, corresponding to the zenith-angle range $\cos\theta_\mu \in [0.2, 0.3]$, where residual cosmic-ray muons are present. The quantities $E_{\nu,k}$ and $\cos\theta_{z,k}$ are the true energy and zenith-angle cosine of the parent neutrino, while $E_{\mathrm{vis},k}$ denotes the total visible energy deposited by detectable final-state particles in the liquid-nitrogen volume, and $\cos\theta_{\mu,k}$ is the zenith-angle cosine of the muon at the entrance to the liquid-nitrogen volume. The indicator function $I_{ij}$ equals 1 if the event falls into the $i$-th zenith-angle bin and the $j$-th visible-energy bin, and 0 otherwise. In this framework, $P_{\mu\mu}$ and $P_{ee}$ denote the $\nu_\mu$ and $\nu_e$ survival probabilities, respectively, while $P_{e\mu}$ and $P_{\mu e}$ denote the $\nu_e\rightarrow\nu_\mu$ and $\nu_\mu\rightarrow\nu_e$ transition probabilities. Contributions from $\bar{\nu}_{\mu}$ and $\bar{\nu}_{e}$ are estimated in the same manner. The oscillation probabilities are thus evaluated using the neutrino truth variables, while the final event distributions used in the sensitivity analysis are stored in binned form in terms of the observables $E_{edep}$ and $\cos\theta_{\mu}$.

For a fixed set of oscillation parameters, the expected muon event sample is accumulated in bins of visible energy and zenith-angle. Table~\ref{tab3} summarizes the scan ranges of the oscillation parameters, and Table~\ref{tab4} gives the visible-energy binning scheme used in the analysis. The variable cos$\theta_{\mu}$ is binned from -1 to 0.3 in steps of 0.1, resulting in 13 bins. The visible energy is divided into 23 bins, which is sufficient to retain the main spectral information in the present ideal-response analysis, while further subdivision yields only a limited additional improvement.

\begin{table}[!htb]
\centering\caption{Oscillation parameter setting.}
\label{tab3}
\begin{tabular*}{8cm} {@{\extracolsep{\fill} } c  c c  c}
\toprule
Parameter  & Input value & Range & Step\\
\midrule
sin$^{2}$$\theta$$_{23}$          & 0.5  & 0.2 - 0.8   & 0.01     \\
sin$^{2}$$\theta$$_{12}$           & 0.304   & Fixed   & -      \\
sin$^{2}$$\theta$$_{13}$            &0.022 & 0.018 - 0.028   & -     \\
$\Delta$m$^{2}$$_{21}$              & 7.5×10$^{-5}$   & Fixed   & -    \\
$\Delta$m$^{2}$$_{32}$               &2.4×10$^{-3}$   & (1 - 4)×10$^{-3}$    & 5×10$^{-5}$  \\
$\delta$$_{CP}$ & 0  & Fixed  & -\\
\bottomrule
\end{tabular*}
\end{table}

\begin{table}[!htb]
\centering\caption{Visible energy bin division.}
\label{tab4}
\begin{tabular*}{8cm} {@{\extracolsep{\fill} } c  c  c}
\toprule
Energy range (GeV)  & No. of bins & Range (GeV)\\
\midrule
0.1-0.5          & 8   & 0.05       \\
0.5-1            & 5   & 0.1       \\
1-2              & 5   & 0.2    \\
2-2.5              & 3   & 0.167     \\
2.5-2.8              & 1   & 0.3  \\
2.8-10             & 1   & 7.2  \\

\bottomrule
\end{tabular*}
\end{table}

\begin{figure}[!htb]
\includegraphics
  [width=1\hsize]
  {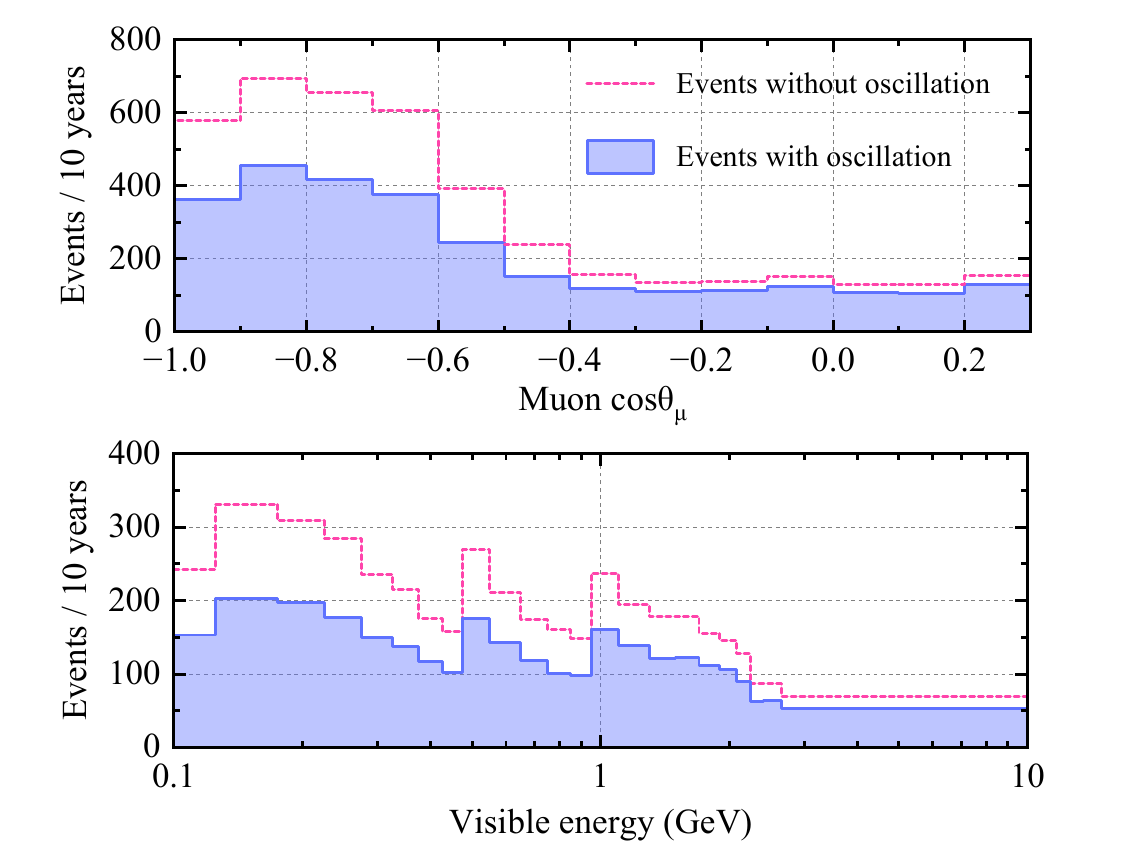}
\caption{Comparison of the total muon zenith-angle cosine and visible-energy distributions before and after oscillations for the 1725~m$^{3}$ liquid-nitrogen cryostat, before applying the cuts. The underlying Geant4 sample corresponds to a 100 year exposure, and the distributions shown are normalized to a 10 year exposure. The oscillation parameters are sin$^{2}$$\theta$$_{23}$ = 0.5, sin$^{2}$$\theta$$_{12}$ = 0.304, sin$^{2}$$\theta$$_{13}$ = 0.022, $\Delta$m$^{2}$$_{21}$ = 7.5×10$^{-5}$ eV$^{2}$, $\Delta$m$^{2}$$_{32}$ = 2.4×10$^{-3}$ eV$^{2}$, and $\delta$$_{CP}$ = 0. }
\label{fig15}
\end{figure}

Figure~\ref{fig15} compares the event distributions before and after oscillations for the 1725~m$^{3}$ liquid-nitrogen cryostat, prior to applying the geometry-based cuts. Muons arising from pion decays induced by $\nu_{e}/\bar{\nu}_{e}$ interactions are included as an oscillation-dependent background, while the residual cosmic-ray muons are added directly to the 13th angular bin. The underlying Geant4 sample corresponds to a 100 year exposure, while the distributions shown in the figure are normalized to a 10 year exposure. The oscillation parameters are sin$^{2}$$\theta$$_{23}$ = 0.5, sin$^{2}$$\theta$$_{12}$ = 0.304, sin$^{2}$$\theta$$_{13}$ = 0.022, $\Delta$m$^{2}$$_{21}$ = 7.5×10$^{-5}$ eV$^{2}$, $\Delta$m$^{2}$$_{32}$ = 2.4×10$^{-3}$ eV$^{2}$, and $\delta$$_{CP}$ = 0. The upper panel shows the one-dimensional $\cos\theta_{\mu}$ distribution of muons before and after oscillations. The oscillation effect is more pronounced at smaller values of $\cos\theta_{\mu}$, corresponding to more upward-going trajectories, and becomes weaker for $\cos\theta_{\mu}$ $\gtrsim$ -0.2. The lower panel shows the corresponding one-dimensional distribution in visible energy. After binning in visible energy, the difference between the spectra before and after oscillations becomes small at high visible energies.

Assuming that the muon event distribution obtained for sin$^{2}$$\theta$$_{23}$ = 0.5, sin$^{2}$$\theta$$_{12}$ = 0.304, sin$^{2}$$\theta$$_{13}$ = 0.022, $\Delta$m$^{2}$$_{21}$ = 7.5×10$^{-5}$ eV$^{2}$, $\Delta$m$^{2}$$_{32}$ = 2.4×10$^{-3}$ eV$^{2}$, and $\delta$$_{CP}$ = 0 is taken as the benchmark (“observed”) data set, we quantify the expected deviation for alternative oscillation parameter choices using a Poisson-likelihood $\chi^{2}$ \cite{bib:45,bib:46}, as defined in Eq.~\eqref{eq:chi2_formula}. Let \(N_{ij}^{\mathrm{obs}}\) denote the benchmark total number of events in the \(i\)-th muon-zenith-angle bin and the \(j\)-th visible-energy bin, including the nominal signal and background contributions, and let \(N_{ij}^{\mathrm{th}}\) denote the corresponding prediction for a trial set of oscillation parameters. Systematic uncertainties are incorporated through pull parameters $\varepsilon_k$, with \(\sigma_{ij}^{k}\) specifying the fractional variation induced by the \(k\)-th systematic effect. The \(\nu_e/\bar{\nu}_e\) background is included through its simulated two-dimensional distribution \(B_{ij}^{\nu_e}\), with a normalization uncertainty of \(\sigma_{\nu_e}=6\%\) described by the pull parameter \(\epsilon_{\nu_e}\). The six common systematic uncertainties are assumed to be correlated between the \(\nu_\mu/\bar{\nu}_\mu\) sample and the \(\nu_e/\bar{\nu}_e\) background and are applied through the same bin-dependent multiplicative factor. For the residual cosmic-ray background, \(B_i^{\rm CR}\) denotes the total contribution in the \(i\)-th angular bin after detector-response smearing. Since its visible-energy distribution is unavailable, it is conservatively distributed uniformly among the 23 visible-energy bins, with its normalization uncertainty described by \(\epsilon_{\rm CR}\) and \(\sigma_{\rm CR}\).
A Gaussian prior is imposed on $\sin^{2}\theta_{13}$, with $\sigma_{\sin^{2}\theta_{13}}$ scanned over its allowed range, and the value that minimizes the total $\chi^2$ is adopted.

The total $\chi^{2}$ is minimized with respect to all pull parameters. Finally, we define $\Delta\chi^{2}$ (Eq.~\eqref{eq:delta_chi2}) as the difference between the $\chi^{2}$ at a given test point and that at the assumed true point. This quantity is used to derive the projected sensitivity of the 1725 m$^{3}$ liquid-nitrogen cryostat to the oscillation parameters.

\begin{equation}
\begin{aligned}
\chi^2 = \sum_{i,j} 2 &\left[ N_{ij}^{\text{th}} - N_{ij}^{\text{obs}} - N_{ij}^{\text{obs}} \ln \left( \frac{N_{ij}^{\text{th}}}{N_{ij}^{\text{obs}}} \right)\right] + \sum_{k=1}^{6} \epsilon_k^2  \\
&+ \epsilon_{\nu_e}^2  + \epsilon_{\text{CR}}^2 
+ \left( \frac{\sin^2 2\theta_{13}^{\text{(true)}} - \sin^2 2\theta_{13}}{\sigma_{\sin^2 2\theta_{13}}} \right)^2, \\
N_{ij}^{\text{th}} &= \left[ N_{ij}^{\nu_\mu} + (1 + \sigma_{\nu_e}\epsilon_{\nu_e}) B_{ij}^{\nu_{e}} \right] \left( 1 + \sum_{k=1}^{6} \sigma_{ij}^k \epsilon_k \right) \\
&\quad + \delta_i^{\text{CR}} \left( \frac{(1+\sigma_{\text{CR}}\epsilon_{\text{CR}})B_{i}^{\rm CR}}{23} \right),
\end{aligned}
\label{eq:chi2_formula}
\end{equation}

\begin{equation}
\Delta\chi^2 = \chi^2(\mathrm{par}) - \chi^2(\mathrm{min}).
\label{eq:delta_chi2}
\end{equation}

\subsection{Systematic uncertainty}
As summarized in Table~\ref{tab5}, six sources of systematic uncertainty are included in the $\chi^{2}$ analysis and are classified into normalization-type and shape/scale-type nuisance parameters. The normalization-type terms are: (k=1) atmospheric neutrino flux normalization (25.6\%), (k=2) an effective 10\% uncertainty on the overall neutrino interaction rate, and (k=6) a 1.7\% effective uncertainty associated with the Geant4 modeling of rock-muon transport and acceptance. The shape/scale-type terms are: (k=3) a 5\% effective uncertainty on the shape of the muon zenith-angle distribution, (k=4) a 5\% effective uncertainty on the shape of the visible-energy spectrum, and (k=5) a 5\% uncertainty on the visible-energy scale. Detector-related effects are treated as effective nuisance parameters that capture residual distortions in the observables, rather than as a full detector-response matrix.

The 25.6\% normalization uncertainty on the atmospheric neutrino flux is obtained by adding in quadrature two contributions. The first is a conservative 25\% baseline uncertainty assigned to the adopted atmospheric neutrino flux model, following the uncertainty scale discussed in the Honda calculation \cite{bib:35} for the low-energy atmospheric neutrino flux. The second is a 5.6\% uncertainty introduced by the geomagnetic-latitude interpolation/extrapolation used to estimate the CJPL flux. The uncertainties associated with hadronic interactions and the primary cosmic-ray spectrum are not introduced again as independent terms, since they are already absorbed into the adopted baseline flux-model uncertainty. The 10\% interaction-rate uncertainty is treated as an effective benchmark normalization in this exploratory sensitivity study. This choice is consistent with the typical scale of neutrino-nucleus cross-section normalization uncertainties adopted in oscillation analyses, where values at the level of about 10–15\% are commonly used for different interaction categories \cite{bib:47}, especially when a full energy-dependent covariance treatment is not available. The 1.7\% Geant4-related term accounts for residual normalization uncertainties in the transport and acceptance of rock-induced muons. To account for possible residual distortions in the observables, we assign a 
5\% zenith-angle shape (tilt) uncertainty, a 5\% spectral tilt uncertainty 
to the visible-energy distribution, and a 5\% scale uncertainty to the 
visible energy. The muon zenith-angle shape uncertainty is modeled as a linear tilt:
\begin{equation}
\sigma_{i}^{\rm zenith}=5\% × \cos\theta_{\mu,i},
\end{equation}
where $i$ denotes the muon zenith-angle bin. The visible-energy spectral 
uncertainty is described by an energy-dependent tilt term:
\begin{equation}
\sigma_{j}^{E}=5\% ×\ln\frac{E_j}{E_0},
\end{equation}
where $E_j$ is the visible-energy bin center and $E_0$ is the reference 
energy, taken as $E_0=1~\mathrm{GeV}$ in this analysis. The coefficient 5\% 
corresponds to the $1\sigma$ uncertainty of the tilt parameter. The 
visible-energy scale uncertainty is implemented as a uniform 5\% shift of 
the visible-energy scale.

The residual cosmic-ray muon background in the extended angular region $0<\cos\theta_\mu<0.3$ is estimated separately. For a 10-year exposure, about 33 residual cosmic-ray muon events are expected in this region, compared with about 110 neutrino-induced events. The background is mainly concentrated near the upper boundary, approximately $0.26<\cos\theta_\mu<0.3$. Considering the finite detector angular resolution, the reconstructed directions of these events are smeared, causing a fraction of them to migrate into neighboring angular bins. The resulting residual cosmic-ray contribution is included in the oscillation analysis as described in Eq.~\eqref{eq:chi2_formula}.

\begin{widetext}
\begin{table*}[!htb]
\centering
\caption{Summary of the systematic uncertainties included in the oscillation analysis.}
\label{tab5}
\begin{tabular*}{17.5cm} {@{\extracolsep{\fill} }l c c c c}
\hline
$k$ & Source & Value & Type & Description \\
\hline
1 & Atmospheric neutrino flux normalization & 25.6\% & Norm. & Baseline model + latitude interpolation \\
2 & Overall neutrino interaction rate & \(10\%\) & Norm. & Effective benchmark uncertainty \\
3 & Muon zenith-angle distribution shape & 5\% & Shape & Angular-distribution distortion \\
4 & Visible-energy spectrum shape & 5\% & Shape & Visible-spectrum distortion \\
5 & Visible-energy scale & 5\% & Scale & Visible-energy scale shift \\
6 & Geant4 rock-muon transport/acceptance & \(1.7\%\) & Norm. & Effective transport uncertainty \\ 
\bottomrule
\end{tabular*}
\end{table*}
\end{widetext}

\begin{figure*}[!htb]
\includegraphics
  [width=0.9\hsize]
  {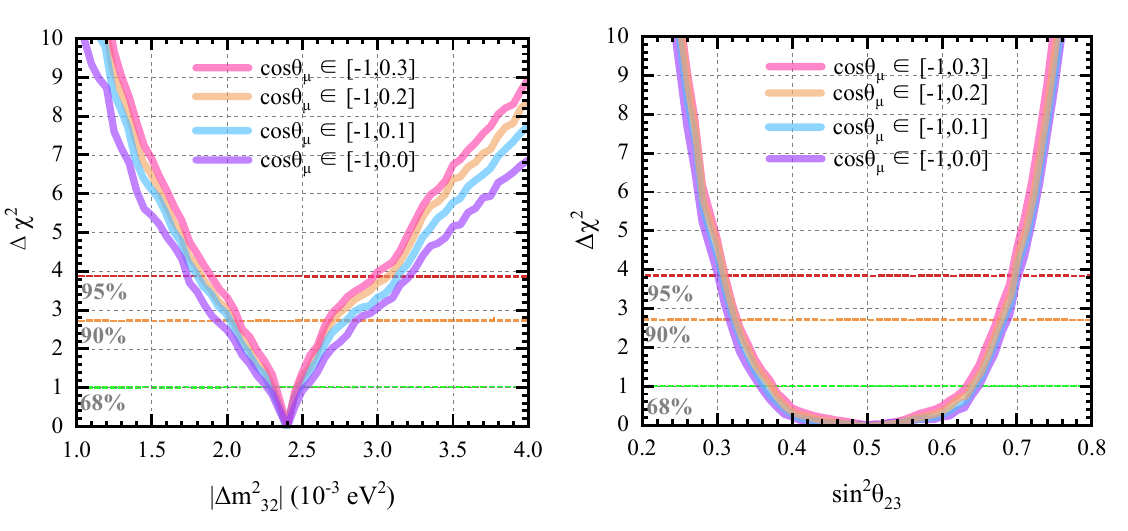}
\caption{Differences in $\Delta$$\chi$$^{2}$ obtained with different zenith-angle ranges after applying the geometry-based cuts. The left panel is evaluated with sin$^{2}$$\theta$$_{23}$ = 0.5 fixed, and the right panel with $\Delta$m$^{2}$$_{32}$ = 2.4×10$^{-3}$ fixed.}
\label{fig16}
\end{figure*}

\begin{figure}[!htb]
\includegraphics
  [width=1.1\hsize]
  {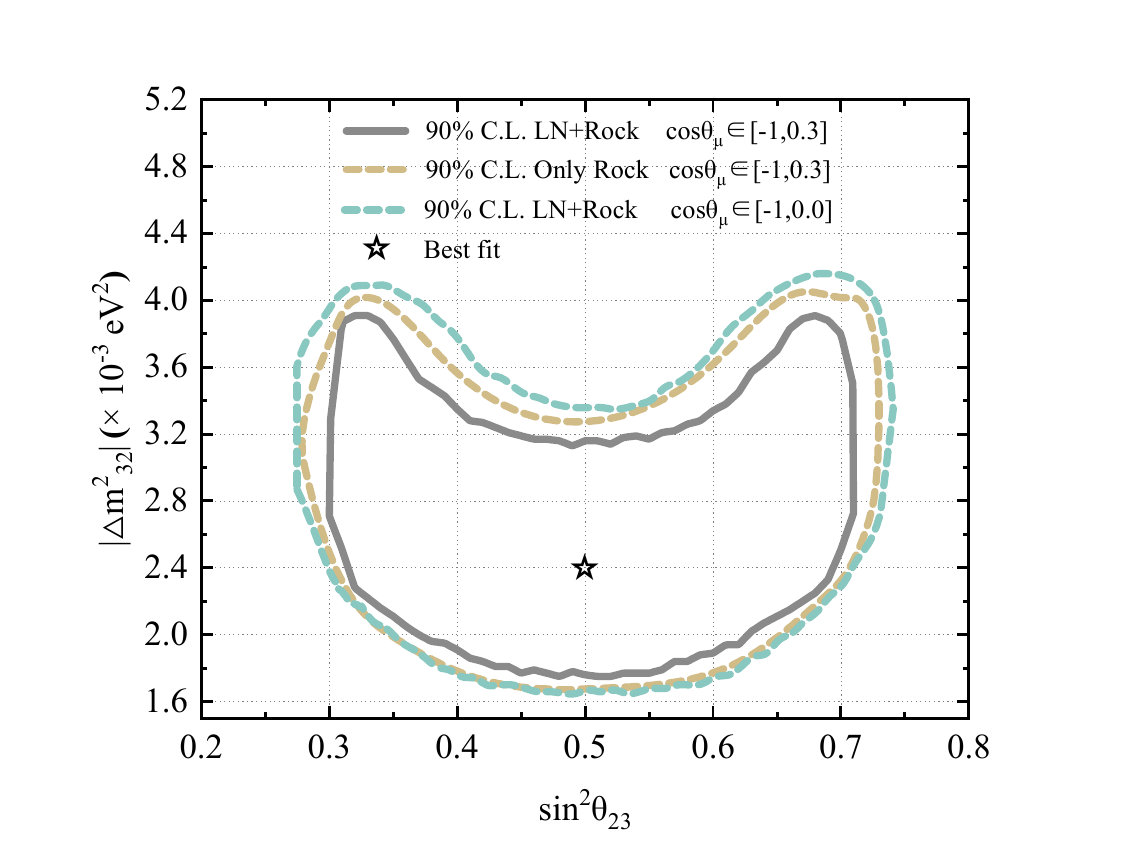}
\caption{The $\Delta$m$^{2}$$_{32}$–sin$^{2}$$\theta$$_{23}$ two-dimensional plane for a $1725~\mathrm{m^{3}}$ liquid-nitrogen cryostat with a 10-year exposure after applying the geometry-based cuts. The 90\% C.L. allowed regions are shown as: the gray solid curve for the combined sample (LN+Rock) with cos$\theta_{\mu}$ $\in$ [-1, 0.3], the brown dashed curve for the rock-induced sample only with cos$\theta_{\mu}$ $\in$ [-1, 0.3], and the cyan dashed curve for the combined sample (LN+Rock) with cos$\theta_{\mu}$ $\in$ [-1, 0].}
\label{fig17}
\end{figure}

\section{Results}\label{sec:6}

After the geometry-based selection, a 10-year exposure at CJPL yields about 2485 rock-induced muon events and 449 events in the liquid nitrogen, which are used for the final oscillation sensitivity analysis. To account for finite detector performance, Gaussian smearing is applied to the Geant4-level observables before constructing the reconstructed event distributions. We assume the reconstruction of muon tracks and visible energy to be comparable to that of established Cherenkov detectors. Based on the performance reported in previous atmospheric neutrino experiments \cite{bib:48}, we adopt a muon angular resolution of $2^\circ$ and a relative visible-energy resolution of 3\% ($\sigma_E = 3\% \times E_{\mathrm{vis}}$). Both energy and angular smearing are applied consistently to all candidate events and background components. The angular smearing also affects the residual cosmic-ray background: after applying the $2^\circ$ resolution, the number of cosmic-ray events in the range $0.2<\cos\theta_\mu<0.3$ increases from 33 to about 42 due to migration from neighboring bins, while no cosmic-ray events appear in the region $0<\cos\theta_\mu<0.2$. The updated cosmic-ray background count of 42 events is used in the subsequent oscillation sensitivity analysis.

Fig.~\ref{fig16} compares the $\chi^{2}$ results obtained with different zenith-angle ranges, while the visible-energy binning is kept fixed. The left and right panels show one-dimensional scans of $\Delta$m$^{2}_{32}$ and $\sin^{2}\theta_{23}$, respectively, fixing the other parameter to $\sin^{2}\theta_{23}=0.5$ or $\Delta$m$^{2}_{32}$ = 2.4$\times10^{-3}~\mathrm{eV}^{2}$. Extending the usable zenith-angle range step by step from $\cos\theta_{\mu}\in[-1,0]$ to $\cos\theta_{\mu}\in[-1,0.3]$ yields a clear improvement in the $\Delta$m$^{2}_{32}$ constraint, whereas the gain for $\sin^{2}\theta_{23}$ is more modest. The additional events in the region $\cos\theta_{\mu}\in[0,0.3]$ correspond to relatively short baselines and therefore small $L/E$, where the oscillation pattern is weak or only beginning to develop. Including this region broadens the phase coverage and helps anchor the onset of the oscillation pattern, thereby improving the determination of the oscillation frequency governed primarily by $\Delta$m$^{2}_{32}$. By contrast, the improvement in $\sin^{2}\theta_{23}$ is more limited, because the weak-oscillation regime carries less information on the oscillation amplitude. To further examine the robustness of these results against possible energy- and angle-dependent shape uncertainties, we define a relative precision measure as
\begin{equation}
\mathrm{Precision}=\frac{P_{1}-P_{2}}{P_{1}+P_{2}},
\end{equation}
where \(P_{1}\) and \(P_{2}\) are the upper and lower bounds of the scanned parameter at a given confidence level. This quantity is used only to compare the relative changes induced by different systematic-uncertainty assumptions, and should not be interpreted as a direct measure of detector performance.

Table~\ref{tab6} and Table~\ref{tab7} summarize the resulting precision for $\Delta$m$^{2}_{32}$ and sin$^{2}\theta_{23}$ under different shape/scale uncertainty settings. The results show that increasing the adopted shape and scale uncertainties leads only to a moderate degradation of the parameter precision. The relative improvement obtained by extending the zenith-angle range is therefore stable against the representative systematic distortions considered here.

\begin{table}[!htbp]
\centering
\caption{Effect of different shape and scale systematic uncertainties on the projected precision of $\Delta$m$^{2}_{32}$.}
\label{tab6}
\begin{tabular}{cccc}
\hline
 & \multicolumn{3}{c}{Precision} \\
\cline{2-4}
Shape \& scale uncertainty & 68\% C.L. & 90\% C.L. & 95\% C.L. \\
\hline
0\%  & 2.7\% & 12.0\% & 20.8\% \\
5\%  & 3.0\% & 12.6\% & 22.4\% \\
10\% & 3.3\% & 13.4\% & 25.1\% \\
\hline
\end{tabular}
\end{table}

\begin{table}[!htbp]
\centering
\caption{Effect of different shape and scale systematic uncertainties on the projected precision of sin$^{2}\theta_{23}$.}
\label{tab7}
\begin{tabular}{cccc}
\hline
 & \multicolumn{3}{c}{Precision} \\
\cline{2-4}
Shape \& scale uncertainty & 68\% C.L. & 90\% C.L. & 95\% C.L. \\
\hline
0\%  & 23.6\% & 33.7\% & 37.2\% \\
5\%  & 25.5\% & 35.0\% & 38.1\% \\
10\% & 25.4\% & 35.3\% & 38.7\% \\
\hline
\end{tabular}
\end{table}

The two-dimensional $\Delta$m$^{2}_{32}$–sin$^{2}\theta_{23}$ plane is shown in Fig.~\ref{fig17}. The figure compares the 90\% C.L. allowed regions for three cases after applying the geometry-based event selections: the combined sample (LN+Rock) with cos$\theta_{\mu}$ $\in$ [-1, 0.3], the rock-induced sample with cos$\theta_{\mu}$ $\in$ [-1, 0.3], and the combined sample (LN+Rock) with cos$\theta_{\mu}$ $\in$ [-1, 0]. Extending the usable zenith-angle range from cos$\theta_{\mu}$ $\in$ [-1, 0] to cos$\theta_{\mu}$ $\in$ [-1, 0.3] leads to a clear contraction of the allowed region, demonstrating an overall improvement in the joint constraint on the oscillation parameters. Consistent with the one-dimensional results, the improvement is more pronounced for $\Delta$m$^{2}$$_{32}$, whereas the reduction in the projected uncertainty of sin$^{2}$$\theta$$_{23}$ around the best-fit $\Delta$m$^{2}$$_{32}$ remains relatively modest. Nevertheless, the contraction of the two-dimensional confidence region away from the best-fit point indicates that the extended zenith-angle coverage also improves the joint determination of $\Delta$m$^{2}$$_{32}$ and sin$^{2}$$\theta$$_{23}$ by reducing the allowed parameter space. For the extended zenith-angle range, the combined sample provides a tighter constraint than the rock-induced sample alone, indicating that the liquid-nitrogen-contained events, although limited in number, still contribute positively to the overall sensitivity. Owing to the limited LN-only statistics, no stable closed 90\% C.L. contour is obtained within the present scan range, and the LN-only result is therefore not shown separately in Fig.~\ref{fig17}.

\section{Summary and discussion}

In this work, we evaluate the projected sensitivity of a 1725~m$^{3}$ liquid-nitrogen-based scenario under CJPL conditions to atmospheric neutrino oscillation parameters for a 10-year exposure. The simulation includes $\nu_{\mu}/\bar{\nu}_{\mu}$ and $\nu_{e}/\bar{\nu}_{e}$ interactions via both CC and NC processes. The oscillation sensitivity analysis is performed using muon event samples, where events induced by the $\nu_{e}/\bar{\nu}_{e}$ component are treated as an oscillation-correlated background, and the residual cosmic-ray muons are included as an additional background. The muon candidate sample comprises $\mu^{\pm}$ produced in neutrino interactions in the surrounding rock, $\mu^{\pm}$ produced within the liquid-nitrogen volume, and $\mu^{\pm}$ originating from $\pi^{\pm}$ decays. For the benchmark oscillation parameters adopted in this study, we obtain a rock-muon flux of $(3.65\pm1.00)\times10^{-13}~\mathrm{cm^{-2}\,s^{-1}\,sr^{-1}}$ and a corresponding muon yield of $(0.13\pm0.034)~\mathrm{day^{-1}}$ in the 1725~m$^{3}$ liquid-nitrogen volume.

Taking advantage of the ultra-low CMBg under CJPL conditions, we extend the usable zenith-angle range in the analysis from cos$\theta_{\mu}$ $\in$ [-1, 0] to cos$\theta_{\mu}$ $\in$ [-1, 0.3]. After applying the geometry-based cuts, $L_{\mathrm{track}}>1~\mathrm{m}$ for rock-induced events and $d_{\mathrm{wall}}>0.3~\mathrm{m}$ for liquid-nitrogen-contained events, 2485 rock-induced events and 449 liquid-nitrogen-contained events are retained for a 10-year exposure. Within this final selected sample, the extended zenith-angle range leads to a visible improvement in both the one-dimensional $\Delta\chi^{2}$ scans and the two-dimensional allowed region in the $\Delta$m$^{2}$$_{32}$–sin$^{2}$$\theta$$_{23}$ plane. The improvement is most pronounced for $\Delta$m$^{2}$$_{32}$, reflecting the enhanced coverage of the oscillation phase provided by the newly included small $L/E$ region. Although the projected improvement in sin$^{2}$$\theta$$_{23}$ at the best-fit point remains relatively modest, the contraction of the two-dimensional confidence region demonstrates an improved joint determination of $\Delta$m$^{2}$$_{32}$ and sin$^{2}$$\theta$$_{23}$ over the explored parameter space.

The present study should be understood as a projected sensitivity analysis based on the current detector assumptions and a simplified treatment of the observables, defined here by the visible energy and the muon zenith-angle cosine. Future developments, including dedicated event reconstruction algorithms, more realistic detector response modeling, and optimized photodetector configurations, will be required to further improve event classification and the reliability of sensitivity evaluations. Within this framework, the results demonstrate that the exceptionally low CMBg at CJPL provides a practical advantage for extending the usable zenith-angle coverage and strengthening the joint constraint on atmospheric neutrino oscillation parameters, with the largest sensitivity gain achieved for $\Delta$m$^{2}_{32}$. More broadly, these results provide a quantitative site-specific reference for future studies of liquid-nitrogen-based detector concepts for atmospheric neutrino physics under deep-underground conditions.

\end{document}